\documentclass[RNAAS,twocolumn]{aastex63}
\usepackage{amsmath}

\newcommand{\jc}[1]{{\textcolor{black}{#1}}}
\newcommand{\cor}[1]{{\textcolor{black}{#1}}}

\newcommand{\nncor}[1]{{\textcolor{black}{#1}}}

\usepackage{lineno}
\shorttitle{Kicking time back in black-hole mergers}
\shortauthors{Araujo et al.}
\graphicspath{{./}{figures/}}
\usepackage[figuresright]{rotating}
\usepackage{subfigure}
\usepackage{booktabs}

\begin{document}

%\title{Kicking down the second-generation nature of the black hole components in GW190521: \\ Ancestral masses, spins and natal recoil}

%and hierarchical-formation viability

\title{Kicking time back in black-hole mergers: \\ Ancestral masses, spins, birth recoils and hierarchical-formation viability of GW190521}

\correspondingauthor{}
\email{araujoalvarezcarlos@gmail.com \\ wyhwong@link.cuhk.edu.hk \\ juan.calderon.bustillo@gmail.com}

\author{Carlos Araújo-Álvarez}
\affiliation{Instituto de Astrofísca de Canarias, 38200 La Laguna, Tenerife, Spain}
\affiliation{Departamento de Astrofísica, Universidad de La Laguna, Tenerife, Spain}
\affiliation{Instituto Galego de F\'{i}sica de Altas Enerx\'{i}as, Universidade de
Santiago de Compostela, 15782 Santiago de Compostela, Galicia, Spain}

\author{Henry W. Y. Wong}
\affiliation{Department of Physics, The Chinese University of Hong Kong, Shatin, N.T., Hong Kong}

\author{Anna Liu}
\affiliation{Department of Physics, The Chinese University of Hong Kong, Shatin, N.T., Hong Kong}

\author{Juan Calder\'{o}n Bustillo}
\affiliation{Instituto Galego de F\'{i}sica de Altas Enerx\'{i}as, Universidade de
Santiago de Compostela, 15782 Santiago de Compostela, Galicia, Spain}
\affiliation{Department of Physics, The Chinese University of Hong Kong, Shatin, N.T., Hong Kong}




\begin{abstract}

Pair-instability supernova (PISN) prevents black-hole formation from stellar collapse within the approximate mass range $M\in [65,130]M_\odot$. However, such black holes may form hierarchically through merging ancestral black holes, whose properties determine those of the ``child'' one: mass, spin, and recoil velocity. Crucially, the child will leave its host environment if its ``birth recoil'' exceeds the corresponding escape velocity, preventing further mergers. We exploit relations between the final recoil and spin of quasi-circular black-hole mergers to obtain posterior probability distributions for the hypothetical ancestral masses, spins and birth recoils of the component black holes of GW190521. To this, we present a Bayesian framework applicable to existing estimates for the components of black-hole merger observations. We consider both the quasi-circular (generically spinning) \textcolor{black}{analysis performed by the LIGO-Virgo-KAGRA collaboration and the eccentric (aligned-spin) one performed by Romero-Shaw et. al. We evaluate the probability $p_{2g}$ that the GW190521 components inferred by these analyses} formed from the merger of stellar-origin black holes and were retained by their environment. For the primary component, which populates the PISN gap, such scenario is strongly suppressed if GW190521 happened in a Globular Cluster \textcolor{black}{with $p_{2g} \sim 10^{-3}$ unless it was quasi-circular and its ancestors had aligned-spins, uncharacteristic of hierarchical formation channels, or small spins, which yields $p_{2g} \simeq 10^{-2}$}. If GW190521 was eccentric, we obtain $p_{2g} \simeq 0.1$ for any host other than an AGN, and zero for a Globular Cluster. If GW190521 was quasi-circular, a Nuclear-Star Cluster origin is possible \textcolor{black}{with $p_{2g} \in (\sim 0.4 \sim ,0.8)$}.

\end{abstract}



\section{Introduction}
\label{SEC: Introduction}

    The way black holes (BHs) form and grow is an open question in astrophysics. While BHs with masses in the approximate range $M\in[5,65]M_\odot$ are well explained through stellar collapse, the formation of more massive ones, like intermediate-mass BHs (with $M\in[10^2,10^5]M_\odot$) or the supermassive BHs populating the center of most galaxies, is still unclear. In this context, and after more than 90 detections of compact mergers \cite{abbott2021gwtc3} by the Advanced LIGO-Virgo network \cite{AdvancedLIGOREF,TheVirgo:2014hva}, now joined by the KAGRA detector, \cite{akutsu2020overview}, gravitational-wave (GW) astronomy represents a powerful tool to shed light on our understanding of BH formation and population \cite{GWTC3-pop}.
    
    While the vast majority of current BBH observations display component BHs consistent with a stellar-collapse origin, some of them deviate from this paradigm. In particular, the event GW190521, originally reported \cite{GW190521D,GW190521I} as a quasi-circular BBH merger with slight evidence for spin-induced orbital precession \jc{(hereafter referred to as ``precession'')}, involves a component BH with a mass ($m_1 = 85 ^{+21}_{-14} M_{\odot}$)\footnote{Expressed as a median value with symmetric $90\%$ credible interval}, squarely fitting within the so-called pair-instability supernova (PISN) gap, which ranges approximately in $M \in [65,130]M_\odot$. According to our current understanding of stellar evolution, BHs in such mass range cannot form from stellar collapse, as the corresponding star would be completely disrupted \cite{Heger:2002by, Woosley2021} \footnote{Moreover, \cite{Farmer:2019jed} proposes that PISN gap may start at even lower values of $45M_\odot$. \cor{Other works, however, propose that the PISN gap could start above $65M\odot$ \cite{Tanikawa2021,2401.17327_Winch}}}. Notably, while orbital precession can be mistaken by eccentricity for GW190521-like events \cite{Bustillo2021_Headon}, alternative studies including orbital eccentricity report a consistent primary mass \cite{Isobel_ecc,Gayathri2022_ecc_natastro,Gamba2022_ecc_natastro}. \jc{The latter is also true for other studies including orbital eccentricity but that find no evidence for it \cite{Toni_ecc}. This indicates that while the analysis of this GW190521 may still be subject to significant unknowns including waveform systematics, conclusions on its primary mass seem to be robust, except if very informative prior distributions on the masses are considered \cite{Fishbach_2020}}. We note that alternative interpretations for GW190521 exist in the literature, including the merger of boson stars \cite{Proca,Observations_proca} or that of dwarf galaxies \cite{Palmese2021}. \jc{Additionally, it has been proposed that BHs populating the PISN gap can be populated by processes such as star-mergers \cite{Costa2022}, convective overshooting \cite{Tanikawa2021} or the collapse of blue supergiants \cite{2401.17327_Winch}.} Despite these options, most hypotheses point that the primary component BH of GW190521 should be a second-generation BH, i.e., formed through a \jc{previous merger of BHs} formed through stellar collapse \cite{GW190521I,Kimball_2021,Anagnostou2022,2209.05766_mahapatra,Gerosa2021_Fischbach_review,Baibhav2021_LIGO_parents_21g_0412} \footnote{See also \cite{GW190521_Nitz,Estells2022_GW190521}}.   

    The above situation has triggered attempts to estimate the masses of the putative ``ancestors'' of GW190521 given the properties of its components, by solving the ``inverse final value'' problem \cite{Barrera2022_masses,Barrera_spins}. This is, deriving the properties of the ``ancestral'' black holes from those of the remnant one. These found that, indeed, its primary BH is consistent with the merger of two BHs with masses below the PISN gap, therefore consistent with a stellar origin. Such studies and the corresponding astrophysical conclusions, however, are limited by three main fundamental aspects. First, they impose that the posterior for the remnant BH, later observed by \jc{GW detectors} as a component of a BBH, is either a delta function at $a_f = 0.7$ (typical of remnants of non-spinning BBHs) or follows a flat distribution within $a_f \in [0,1]$. This does not exploit the rich information encoded in the actual probability distribution measured through the observation of GW190521 \textit{which, as we will show, is crucial for our conclusions}. Second, when considering the ancestral spins, they restrict these to be aligned with the angular momentum, therefore omitting orbital precession \cite{Barrera_spins} \jc{which is highly likely to occur in dynamical environments like globular clusters 
    \cite{Talbot2017,Rodriguez2016}}. Finally, both works omit one of the key aspects of this work, namely \textit{the impact of the gravitational recoil or ``birth kick''} inherited by BHs born from BBHs. Such information is vital, as the kick determines whether the remnant BH can remain within its environment and undergo the subsequent merger eventually observed by our detectors.
    
    The asymmetric emission of GWs by BBHs produces a net emission of linear momentum \cite{Thorne:1980ru}. This imparts a gravitational recoil, or ``kick" \cite{Gonzalez:2006md}, to the remnant BH which can reach thousands of km/s in the most extreme cases \cite{Brgmann2008,Sperhake2011} . If the kick exceeds the escape velocity $v_{\text{esc}}$ of the host environment, the BH will abandon it, preventing its participation in subsequent mergers. Therefore, retaining a second-generation BH within its host environment requires the latter to have an escape velocity that exceeds the BH kick. To date, kick estimates in existing GW events have focused on that inherited by the final BH \cite{CalderonBustillo:2018zuq,Varma2020_kickstudy, Vijay_GWKick,Measured_kick_magnitude_GW190814,GW190412_recoil}. In contrast, in this work, we perform for the first time estimates of the (putative) birth kicks of the component BHs observed in BBH observations, with the goal of assessing their viability as products of previous mergers of stellar-origin BHs as a function of the properties of the host environment. \\ %This can repeat, producing increasingly more massive black holes at each step in process called hierarchical mergers (HMs). HMs are one of the considerated BH alternative formation channels (ref), though its efficiency and predominance is still unclear (ref). 

\begin{figure*}
        \includegraphics[width=0.33\textwidth]{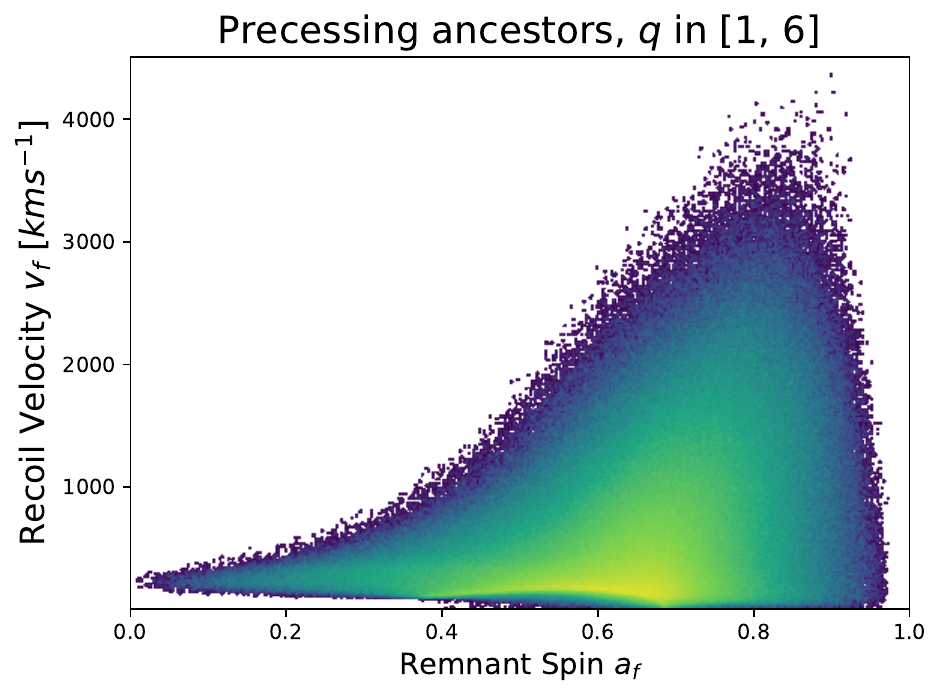}
        \includegraphics[width=0.33\textwidth]{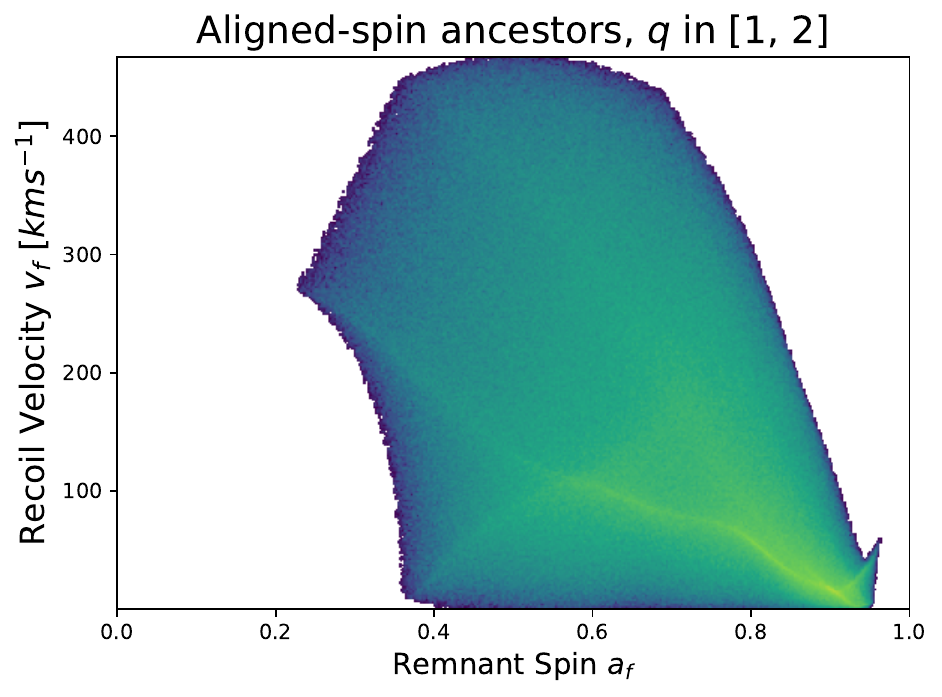}
         \includegraphics[width=0.33\textwidth]{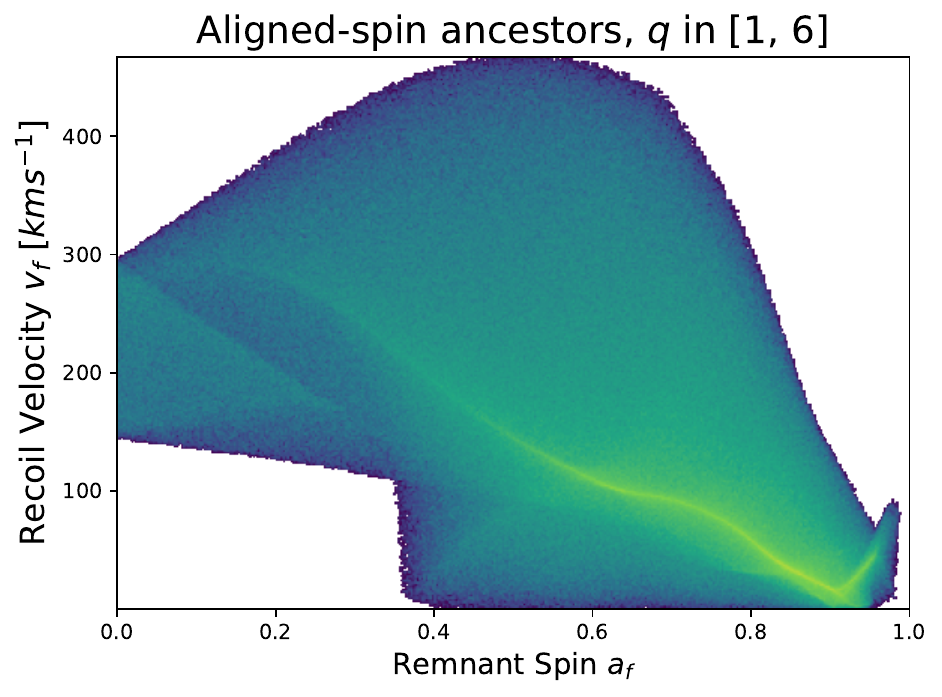}
        \includegraphics[width=\textwidth]{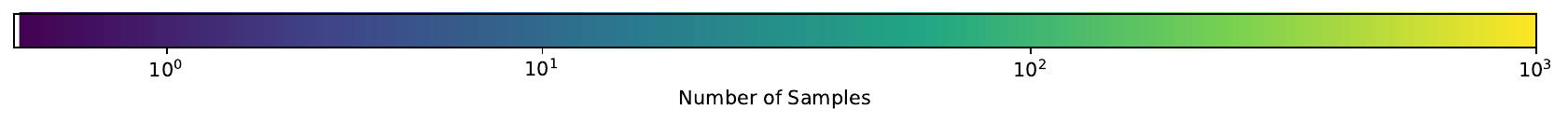}
        \caption{\textbf{Relation between remnant spin and kick in quasi-circular black-hole mergers}. The left panel assumes an isotropic spin prior allowing for orbital precession and mass ratios $q\leq 6$ while the central and right ones are restricted to aligned-spins, parallel to the orbital angular momentum of the binary, and mass ratios $q\leq 2$ and $q\leq 6$ respectively. The color code denotes the  probability density. }
        \label{FIG: Prior_distribution}
    \end{figure*}

    Making use of existing full parameter estimates for the components of GW190521, we infer the masses and spins of their putative ancestor BHs, together with the corresponding birth kicks. \textcolor{black}{As input analyses, we consider the original analysis performed by the LIGO-Virgo-KAGRA collaboration (LVK), which assumes that GW190521 was a quasi-circular merger \cite{GW190521D}, and that by \cite{Isobel_ecc}, which considers an eccentric non-precessing merger. In addition, we consider three different proposal distributions for the parameters of the ancestral BBHs motivated by those expected from dynamical and isolated formation channels}. Importantly, to estimate birth kick, we exploit the tight relations between the final kick and final spin of quasi-circular BBH mergers shown in Fig. \ref{FIG: Prior_distribution}, which we will later discuss. Finally, we assess the viability of the GW190521 component BHs as the result of the merger of stellar-origin BHs. To this, we compute the probability that \textcolor{black}{the GW190521 components admit ancestors within our proposal distributions, with both having masses $(m_{i1},m_{i2})<65 M_\odot$ at the same time that the birth kick $v_f$ induced in the remnant black hole is lower \jc{than} the escape velocity $v_{esc}$ of a given host environments.}

\section{Method}
\label{SEC:Methodology}

   % \subsection{GW190521 parameter estimates}

        %Our \jc{starting} point are existing posterior probability distributions for the masses $(m_{1},m_{2})$ and spin-magnitudes $(a_{1},a_{2})$ of the component BHs of GW190521. We use posteriors obtained both by the LVK \cite{GW190521D}  (assuming a quasi-circular merger scenario) and by Romero-Shaw et. al. \cite{Isobel_ecc} (assuming an eccentric merger). Next, we construct prior probability distributions for the masses and spins of the ancestral BHs $p(\theta_{\rm anc} \equiv \{m_{i1},m_{i2},\vec{a_{i1}},\vec{a_{i2}}\})$, which we will later discuss. From these, we obtain the induced prior distributions $p(m_f,a_f,k_f)$ for the final BH mass $M_f$ and final spin $a_f$ (which we will use to connect the GW190521 components with its ancestors), together with the remnant recoil velocity $v_f$. We obtain these using the surrogate model for the final BH properties $\texttt{NRSur7dq4Remnant}$ \cite{NRSur7dq4,Varma2019_surrogate_remnant}.\\

\subsection{Estimating the components of GW190521}

\textcolor{black}{Consider a stretch of gravitational-wave strain data $d(t)$ and a waveform model $h(t;\theta_{\rm BBH})$ for the gravitational-wave emission of a black-hole merger with parameters $\theta_{\rm BBH}$. The posterior probability $p(\theta_{\rm BBH}|d)$ for $\theta_{\rm BBH}$  is given by}

 \begin{equation}
     p(\theta_{\rm BBH}|d) = \frac{p(d|\theta_{\rm BBH}) p(\theta_{\rm BBH})}{p(d)}.
 \end{equation}

\textcolor{black}{In the above expression,  $p(\theta_{\rm BBH})$ denotes the prior probability for the source parameters $\theta_{\rm BBH}$, the term $p(d|\theta_{\rm BBH})$ denotes the probability of the data given the $\theta_{\rm BBH}$ -- known as likelihood -- and the normalization term denotes the Bayesian evidence, given by}

\begin{equation}
     p(d) = \int p(d|\theta_{\rm BBH}) p(\theta_{\rm BBH}) d\theta_{\rm BBH}
\end{equation}

\textcolor{black}{The likelihood is given by the well-known expression \cite{Finn1992, Romano2017}}

 \begin{equation}
   \log  p(d|\theta_{\rm BBH}) \propto \frac{1}{2}(d-h(\theta_{\rm BBH})|d-h(\theta_{\rm BBH})),
 \end{equation}

\textcolor{black}{where $(a|b)$ denotes the inner product} 

\begin{equation}
(a|b) = 4 \Re \int \frac{\tilde{a}(f)\tilde{b}^{*}(f)}{S_n(f)} df.
\end{equation}

\textcolor{black}{Above $\tilde{a}(f)$ denotes the Fourier transform of $a(t)$ and $*$ denotes complex conjugation. Finally, the term $S_n(f)$ denotes the one-sided spectral density of the gravitational-wave detector.}\\

\subsection{Estimating the ancestor parameters}

\textcolor{black}{Consider the posterior distribution $p(\theta_{\rm BBH}|d)$ for the component parameters of a BBH conditional on the gravitational-wave data $d$. Our aim will be to derive posterior distributions for the putative ancestors of the component black holes of the BBH. Let us, however, first discuss the easier and more common exercise of obtaining posterior distributions for the remnant parameters, which we will then extend.}

\paragraph{The ``trivial'' example of remnant parameters}

\textcolor{black}{Given the posterior distribution for the component parameters of a BBH, General Relativity provides a deterministic relationship for predicting the parameters of the final black hole $\theta_{f}(\theta_{\rm BBH})$. In this work, we will compute the final black hole properties using the $\texttt{NRSur7dq4Remnant}$ model \cite{NRSur7dq4,Varma2019_surrogate_remnant}. With this, the posterior distribution for the remnant parameters $\theta_{f}$ can be constructed as}

 \begin{equation}
     p(\theta_{f}|d) = \int p(\theta_{f}|\theta_{\rm BBH}) p(\theta_{\rm BBH}|d) d\theta_{\rm BBH}
 \end{equation}

\textcolor{black}{In the above, the term $p(\theta_{f}|\theta_{\rm BBH})$ is simply given by} 

  \begin{equation}
     p(\theta_{f}^{'}|\theta_{\rm BBH}) = \delta(\theta_{f}(\theta_{\rm BBH}) - \theta^{'}_{f}),
 \end{equation}

\textcolor{black}{which takes a value of 1 if the remnant properties $\theta_{f}(\theta_{\rm BBH})$ of the BBH match $\theta^{'}_{f}$, and 0 otherwise.}

\paragraph{Ancestral parameters}
\textcolor{black}{We now aim to extend the above methodology to estimate the parameters $\theta_{\rm anc}$ of the putative ancestors of a component black hole of a BBH. In this context, we treat the observed black-hole parameters as ``child'' parameters, denoted by $\theta_{\rm child}$. Our goal is to infer the posterior distribution for the ancestor parameters $p(\theta_{\rm anc}|d)$ given the gravitational-wave data. This is given by}

 \begin{equation}
     p(\theta_{\rm anc}|d) = \int p(\theta_{\rm anc}|\theta_{\rm child}) p(\theta_{\rm child}|d) d\theta_{\rm child}.
 \end{equation}

\textcolor{black}{The function $\theta_{\rm child}(\theta_{\rm anc})$ is however not invertible, since a black hole with parameters \textcolor{black}{$\theta_{\rm child}$} can be the result of multiple combinations of $\theta_{\rm anc}$. Consequently, the term $p(\theta_{\rm anc}|\theta_{\rm child})$ is not uniquely defined, leaving us the freedom propose it. In other words, given a child black hole, we need to propose a probability distribution that determines the relative a-priori probability of its possible ancestors.}\\ 

\textcolor{black}{To proceed, we simply construct this as the conditioning of some probability distribution $p(\theta_{\rm anc})$ for the ancestral parameters to the subspace satisfying $\theta_{\rm child}(\theta_{\rm anc}) = \theta^{*}_{\rm child}$. This is obtained as}:

\begin{equation}
        p(\theta_{\rm anc}|\theta^{*}_{\rm child}) = \frac{p(\theta_{\rm anc}) \cdot \delta(\theta_{\rm child}(\theta_{\rm anc}) - \theta^{*}_{\rm child})}{{\cal{Z}}(\theta^{*}_{\rm child})},
\end{equation}

\textcolor{black}{where the normalization factor ${\cal{Z}}(\theta^{*}_{\rm child})$ is given by}

\begin{equation}
\begin{aligned}
       {\cal{Z}}(\theta^{*}_{\rm child}) = 
      \int p(\theta_{\rm anc}) \cdot \delta(\theta_{\rm child}(\theta_{\rm anc}) - \theta^{*}_{\rm child}) d\theta_{\rm anc}.
\end{aligned}
\end{equation}

\textcolor{black}{We note that the above term denotes the prior probability on the parameters $\theta_{\rm child}$ that would be induced by the distribution $p(\theta_{\rm anc})$ if this was enforced as a prior for the analysis of the gravitational-wave data. We intentionally denote this by ${\cal{Z}}(\theta_{\rm child})$ to avoid confusion with the prior $p(\theta_{child})$ on the component BBH parameters imposed by our input analyses of GW190521, which we inherit.}\\

\textcolor{black}{With this, the posterior distribution for the ancestral parameters is given by:}

\begin{equation}
\begin{aligned}
&p(\theta_{\rm anc}|d) = \\ &  p(\theta_{\rm anc}) \int   \delta(\theta_{\rm child}(\theta_{\rm anc}) - \theta^{*}_{\rm child}) \frac{p(\theta^{*}_{\rm child}|d)}{{\cal{Z}}(\theta^{*}_{\rm child})} d \theta^{*}_{\rm child} 
\end{aligned}
\end{equation}

\textcolor{black}{Above, the posterior distribution $p(\theta_{\rm child}|d)$ is directly obtained from existing analyses of GW190521.}

\paragraph{On the choice of $p(\theta_{\rm anc})$} \textcolor{black}{As we will highlight later, we note that certain choices $p(\theta_{\rm anc})$, regardless their astrophysical motivation, will lead to a situation where no ancestral black holes can be found for a fraction of the component black-holes supported by our input analyses of GW190521. For instance, ancestral binaries restricted to equal masses cannot form remnant black holes with null (and low) spin, which are clearly supported by the posterior distributions for GW190521 that we take as input. Therefore, such a choice of $p(\theta_{\rm anc})$ will imply a reduction of the probability $p(\rm anc | d)$ that the GW190521 components estimated by our input analyses were born from an ancestral merger, which is given by} 

\begin{equation}
\begin{aligned}
    & p(\rm anc | d) = \\ & \int d\theta_{child} p(\theta_{child}|d) \int p(\theta_{child}|\theta_{anc}) \frac{p(\theta_{anc})}{{\cal{Z}}(\theta_{child})} d\theta_{anc}.
\end{aligned}
\end{equation}

\textcolor{black}{The above represents the fraction of the posterior distribution for $\theta_{\rm child}$ for which ancestors can be found within $p(\theta_{\rm anc})$. This will be equal to 1 for broad enough $p(\theta_{\rm anc})$.}\\

\textcolor{black}{We stress that the above does not necessarily imply that restrictive priors on $\theta_{\rm anc}$ provide a bad model for the \textit{observed gravitational-wave data}, if this was analysed under the corresponding induced priors on $\theta_{\rm child}$. In fact, such restrictive ancestral priors may actually yield better models for the observed data if these induce child-priors that favour the highest likelihood regions of the parameter space spanned by $\theta_{\rm child}$. Explicitly checking this, however, requires in principle of the ``forward'' framework proposed by \cite{Mahapatra2024} shortly after the release of this work -- which we discuss next -- as opposed to our ``backwards'' one, where we take existing analyses of GW190521 as input to then reconstruct the history of the estimated component black holes.}

\textcolor{black}{\paragraph{Comparison to other methods} In this ``backwards'' approach, we take as input data posterior distributions for the child black hole parameters obtained through existing analyses and aim to reconstruct the distributions of their ancestors. This method contrasts with the ``forward'' approach of  \cite{Mahapatra2024}, where the authors impose a prior $p(\theta_{\rm anc})$
on the ancestor parameters which in turn induces a prior $p(\theta_{\rm child})$ on the child ones. This induced prior is then inputted in eq. (1) to analyze the gravitational-wave data, leading to new posterior distributions for the child parameters.}

\textcolor{black}{The key difference between the two methods is that our ``backwards'' approach preserves the original prior and posterior distributions for the child black holes obtained from ``standard'' gravitational-wave data analyses, whereas the ``forward'' method does result in different posteriors for these child parameters. In other words, while our method is focused on \textit{reconstructing the ancestor parameters based on the known posteriors of the child black holes}, the forward approach simultaneously seeks to determine both the ancestor and child parameters from the data, starting from a prior on the ancestor population.}

\textcolor{black}{While both methods address valid and well-posed questions, we believe our approach offers at least one advantageous aspect. By preserving the broad parameter priors commonly used in gravitational-wave data analyses for the child black holes, we avoid the potential for overly restrictive priors on the child parameters. In the ``forward'' method, such restrictive priors may be induced by those imposed on the ancestors, possibly preventing the exploration of certain regions of the parameter space for the child black holes. This could limit the range of physically plausible solutions, }\textcolor{black}{although it has the advantage of allowing one to test particular astrophysically motivated distributions for the ancestral black holes.}

\textcolor{black}{\paragraph{Interpretation of $p(\theta_{\rm anc})$ as a prior on the ancestors} Finally, we want to comment on the interpretation of the term $p(\theta_{\rm anc})$ as a prior in our analysis. As is obvious, one cannot freely impose priors on both $p(\theta_{\rm anc})$ and  $p(\theta_{\rm child})$, as the imposition of one automatically induces the other. As mentioned, in our work we do inherit the prior $p(\theta_{\rm child})$ imposed in our input analyses of GW190521 which, induces constraints on that for the ancestors. However, we note that eq. (5) shows that the posterior distributions on the ancestral parameters obtained through our formalism are \textit{proportional} to -- and therefore can be interpreted as --  those that would be obtained in an analysis where one imposes a prior $p(\theta_{\rm anc})$ on the ancestral parameters \textit{and conditions the posterior distributions for $\theta_{anc}$ on those for the child parameters given by our input analyses, instead of conditioning directly on the gravitational-wave data}. Importantly, in such analysis, the likelihood is given by the integral term (i.e., governed by the posterior distribution on the child parameters) and not by the gravitational-wave likelihood directly involving the gravitational-wave data, shown in eq. (3).} %We note, however, that within latter analysis the Bayesian evidence is given by 

        \subsection{Practical implementation}
        
        In practice, we implement the following procedure to obtain samples from the posterior distribution of the ancestral BH parameters:
        \begin{enumerate}
            \item First, we generate $2\times 10^6$ random samples for the ancestral prior $p(\theta_{\rm anc})$, computing the corresponding  $\{m_f,a_f,k_f\}(\theta_{\rm anc})$.
            \item \textcolor{black}{Second, we retrieve the posterior distribution $p(\theta_{\rm child}|d)$ returned by an existing analysis of GW190521. As said before we consider those presented in \cite{GW190521D} and \cite{RomeroShaw2020_ecc_apjl}.} 
            \item For each sample $\{m_i,a_i\}$ of the posterior distribution for GW190521 $p(\theta_{\rm child}|d)$, we draw random samples from $p(\theta_{\rm anc})$ satisfying $\{m_f,a_f\} (\theta_{\rm anc}) = \{m_i\pm \delta m_i,a_i\pm \delta a_i\}$. \textcolor{black}{Again, we stress that, depending on the choice of $p(\theta_{anc})$, this will only be possible for a fraction $p(\rm anc | d)$ of the posterior samples.}
        \end{enumerate}
    
        The usage of non-zero tolerances $\delta m_i$, $\delta a_i$ is motivated by the fact that the discrete nature of our samples for $p(\theta_{\rm anc})$ would naturally prevent us from encountering samples exactly matching the condition imposed in Step 3 above. In our case, we set $(\delta m_i, \delta a_i) = (1,0.05)$, checking that this produces stable results. \\  
    
        Finally, we compute the probability that the component BHs of GW190521 are the product of the merger of stellar-origin \jc{BHs}. To do this, \textcolor{black}{we multiply the probability $p(\rm anc|d)$ that viable ancestors exist given our choice of $p(\theta_{\rm anc})$ by the probability $p(\rm 2g | \rm anc)$ that such ancestors have masses below $65M_\odot$, at the same time that $k_f < v_{esc}$, with $v_{esc}$ denoting the escape velocity of the host environment. This is, we compute}
         \begin{equation}
         \begin{aligned}
           & \textcolor{black}{p_{2g} = p(\rm anc | d}) \cdot p(\rm 2g | d) = \\ & p(\rm anc | d) \cdot p(k_f <v_{esc}, m_{1,2}<65M_\odot | \rm anc)
        \end{aligned}
        \end{equation}
    
        %We compute also the fraction of remnants with a kick smaller than a certain escape velocity $v_{esc}$ and whose progenitors masses were both under 65 $M_{\odot}$. This fraction is given by
    
        %\begin{equation}
        %    \label{EQU: probability distribution}
        %    p(v<v_{esc} | m_{ij}<65) = \sum_{k<l} p(v_{k}|m_{ij}<65)
        %\end{equation}
    
        %where $v_{i}$ is the acquired gravitational recoil, $v_{l}=v_{esc}$ and $m_{ij}$ are the masses of %ancestors,given in solar mass units.

       % We compute the final spin and recoil velocity of a simulated population of $2 \times 10^6$ BBH mergers to establish an uninformative uniform prior using \texttt{SurfinBH} \citep{varma2019high}.

    \subsection{Choice of the ancestral distributions $p(\theta_{\rm anc})$}

        The choice of the ancestral \textcolor{black}{distribution $p(\theta_{\rm anc})$} can have a strong impact on the results, specially in cases where the spins of the component BHs are not well constrained by our input analyses. First, the \textcolor{black}{distribution} for the ancestral masses will \jc{clearly} impact the chances that both of them are below $65M_\odot$. Second, regarding the ancestral spins, it is well known that while precessing BBHs can produce recoils exceeding $3000$ km/s, those for aligned-spin BBHs are capped at around $500$ km/s \cite{Healy2014_alignedspinkick}. Therefore, different assumptions for the hypothetical ancestral masses and spins can strongly impact the hierarchical-origin viability of the observed BHs.\\
        
        In light of the above, we consider three \textcolor{black}{distributions $p(\theta{\rm anc})$} for the ancestral BBHs:\\
        
        a) A ``precessing \textcolor{black}{distribution}'' (PP) where we consider generically spinning BBHs with spins isotropically distributed on the sphere and spin magnitudes uniformly distributed in $a_i \in [0, 0.99]$. We pair this with priors on the masses $m_{i} \in [5, 170]$ $M_{\odot}$, broad enough to encompass \textcolor{black}{the ancestral masses consistent with the components of GW190521}. We constrain the mass ratio to $q\in [1, 6]$ which corresponds to the domain of the remnant model \texttt{NRSur7dq4Remnant}. \\
        
        b) A ``restricted \textcolor{black}{distribution}'' aligned-spin prior (APq2) where we consider binaries with masses uniformly distributed in $m_{i} \in [5, 170] M_{\odot}$  restricted to mass ratios $q\in [1, 2]$. We also impose aligned spins, parallel to the orbital angular momentum of the binary, with magnitudes uniformly distributed in $a_i \in [0, 0.99]$.\\
        %Fig. \ref{FIG: Priors} shows the one-dimensional distributions of the recoil velocity and the final spin of the remnants in our simulations.

        c) An aligned-spin \textcolor{black}{distribution} (APq6), identical to the APq2 in b), but expanded to mass ratios $q\leq 6$.\\

        The three above choices closely relate to the possible formation channels of the ancestral binaries and their host environments \cite{Mapelli2020,Farr:2017uvj}. While isotropic spin distributions are expected in dynamical formation scenarios characteristic of dense environments \cite{Tutukov1993,Belczynski2016}, aligned spins are characteristic of isolated binary formation \cite{Sigurdsson1993,PortegiesZwart2000,Rodriguez2016}. \cor{While both theoretical studies \cite{Fuller2019} and observational data \cite{GWTC3-pop} suggest that first-generation BHs tend to have small spins, we chose spin-magnitude priors spanning the whole $a_i \in [0,1]$ range. \textcolor{black}{Following our previous discussion on restrictive choices for $p(\theta_{\rm anc})$, we want to keep these as agnostic as possible} to avoid missing any potential ancestors of the GW190521 components \textcolor{black}{inferred by our input analyses. In fact, we have checked that choices of $p(\theta_{\rm anc})$ restricting spin magnitudes to both null spins and $a_i < 0.2$ -- more characteristing of stellar-origin black holes -- fail to reproduce between $50\%$ and $80\%$ of the posterior spin distributions of the GW190521 components, therefore yielding values of $p(\rm anc | d)$ between 0.5 and 0.2}. \textcolor{black}{As also stressed before, we note that this does not necessarily imply that ancestral priors following such distributions -- together with those induce on $\theta_{\rm child}$ -- would pose a bad model for the gravitational-wave data. Instead, it means that they are not consistent with a second-generation origin of the black-holes inferred by our input analyses}, \textcolor{black}{and under their corresponding priors.} We have checked, nevertheless, that despite allowing for much smaller final recoils than $p(\theta_{anc})$ distributions allowing for generic spins, such models yield very similar retention probabilities, therefore not strongly modifying our conclusions.}\\
        
        The restriction of the aligned-spin \textcolor{black}{$p(\theta_{\rm anc})$} distribution to mass ratios $q\leq 2$ is motivated by results indicating that BBHs formed through isolated scenarios tend to display such small mass ratios \cite{Belczynski2016,Giacobbo_2018,Mandel_2016}. \cor{We note, however, that in principle, binaries formed in isolation should not lead to repeated mergers. It has been noted, nevertheless, that binaries in Active Galactic Nuclei (AGN) disks may tend to align their spins with the angular momentum of the accretion disk \cite{Natarajan1998,Yang2019}, providing another good motivation to consider an aligned-spin ancestral distribution. In any case, we will use the aligned-spin prior to showcase the strong impact of the induced birth kicks in our study}.\\ 
        
        As we will show, each choice of \textcolor{black}{$p(\theta_{\rm anc})$} leads to significantly different final spin distributions, as well as different kick distributions for a given final spin. As a consequence, on the one hand, this means these scenarios will have different abilities to produce the \textcolor{black}{GW190521 component BHs inferred by our input analyses}. On the other hand, even if both scenarios could form such BHs, the corresponding kicks will widely differ, leading to very different probabilities for the BHs to be retained their environments and undergo subsequent mergers. We note, however, that we have found that the restriction to $q\leq 2$ in our APq2 distribution fails to reproduce most of \textcolor{black}{our input posterior distributions} for the spin magnitudes of GW190521 $a_{1,2}$, \textcolor{black}{leading to low values of $p(\rm anc | d)$}. This is the reason why we consider our distribution c), with mass ratios up to $q\leq 6$. Finally, we note that several works \cite{Fishbach_2020,GW190521_Nitz,koustav_2309.01683} have shown the important role of mass-prior choices in analysing high-mass events. For this reason, we have also considered the above three priors, but using mass-priors uniform in total mass $M$ and mass ratio $q\in[1,6]$. We have found, however, that our results are barely sensitive to this change.
        
\subsection{Bayesian priors on ``child'' parameters inherited from our input analyses}

\textcolor{black}{Here we describe the priors on the child parameters imposed by our two input analyses. We stress that these priors are rather generic, with the goal of encompassing all possible parameter combinations consistent with the GW data and therefore do not represent expectations of actual distributions of astrophysical sources.}

\begin{itemize}
\item \textcolor{black}{Quasi-circular ``LVK'' scenario:  this analysis imposes uniform priors on the red-shifted component masses $m_i^{z} = m_i (1+z)$ of GW190521, with $z$ denoting the redshift,  together with a prior in the luminosity distance.  In addition, it imposes isotropic priors on the spin directions and uniform priors on their magnitudes, ranging in $a_i \in [0,1]$.}
\item \textcolor{black}{Eccentric ``Romero-Shaw'' scenario:  this analysis imposes identical priors on the component masses of GW190521 as the one above. Regarding the spins, these are restricted to be aligned or anti-aligned with the orbital angular momentum, with magnitudes ranging uniformly in $a_i \in [0,1]$.}
\end{itemize}

        %Fig. \ref{FIG: Priors} shows the one-dimensional distributions of the recoil velocity and the final spin of the remnants in our simulations.

\section{Results}
\label{SEC: Results and Discussion}

    \begin{figure*}
        \includegraphics[width=0.5\textwidth]{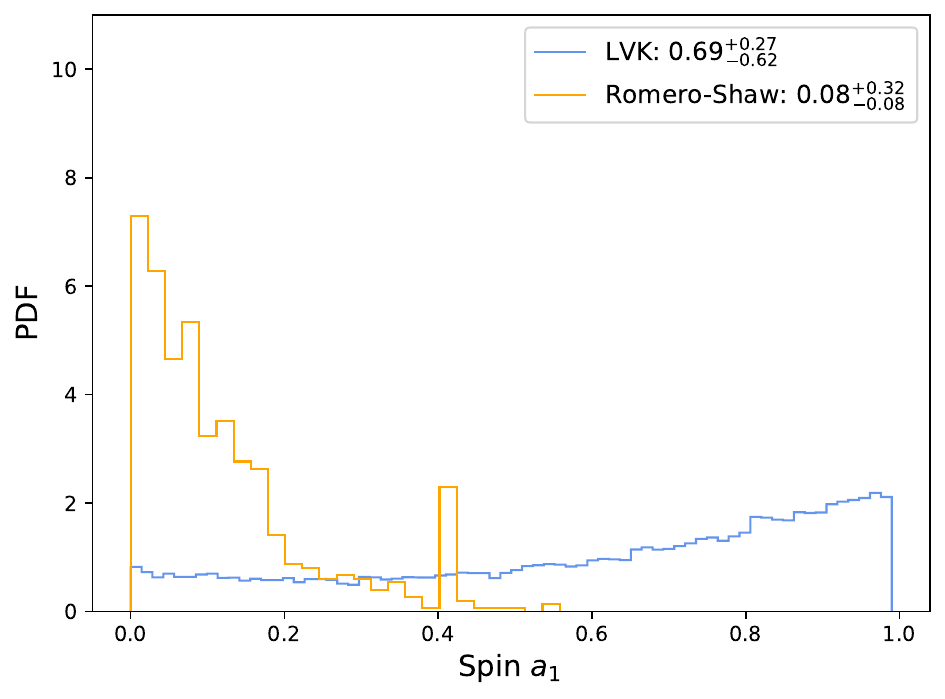}
        \includegraphics[width=0.5\textwidth]{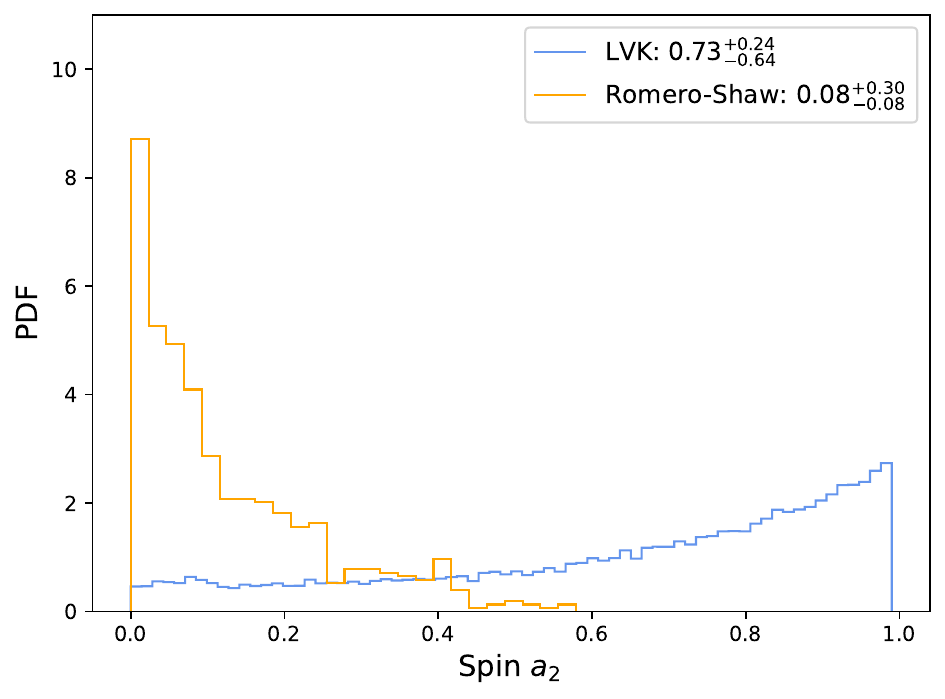}
        \caption{\textbf{Spin magnitudes for the GW190521 component black holes}. The left (right) panel corresponds to the primary (secondary) black hole. Blue distributions correspond to the quasi-circular precessing scenario for GW190521, originally reported by the LVK \cite{GW190521D}, while the orange ones correspond to the eccentric non-precessing scenario described in Romero-Shaw et. al. \cite{Isobel_ecc}.}
        \label{FIG: Spin_posterior}
    \end{figure*}

    \begin{figure*}
        \includegraphics[width=0.33\textwidth]{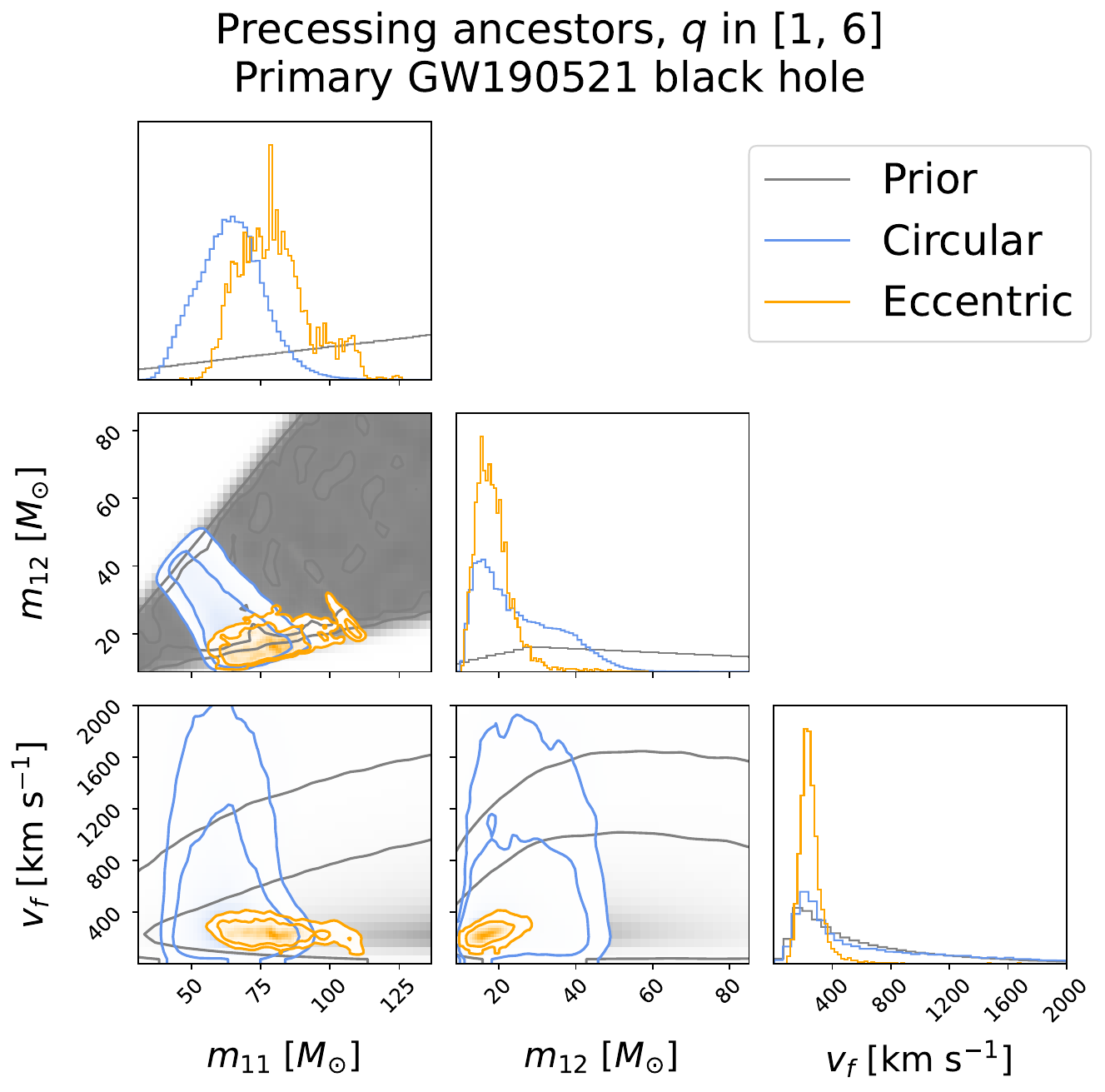}
        \includegraphics[width=0.33\textwidth]{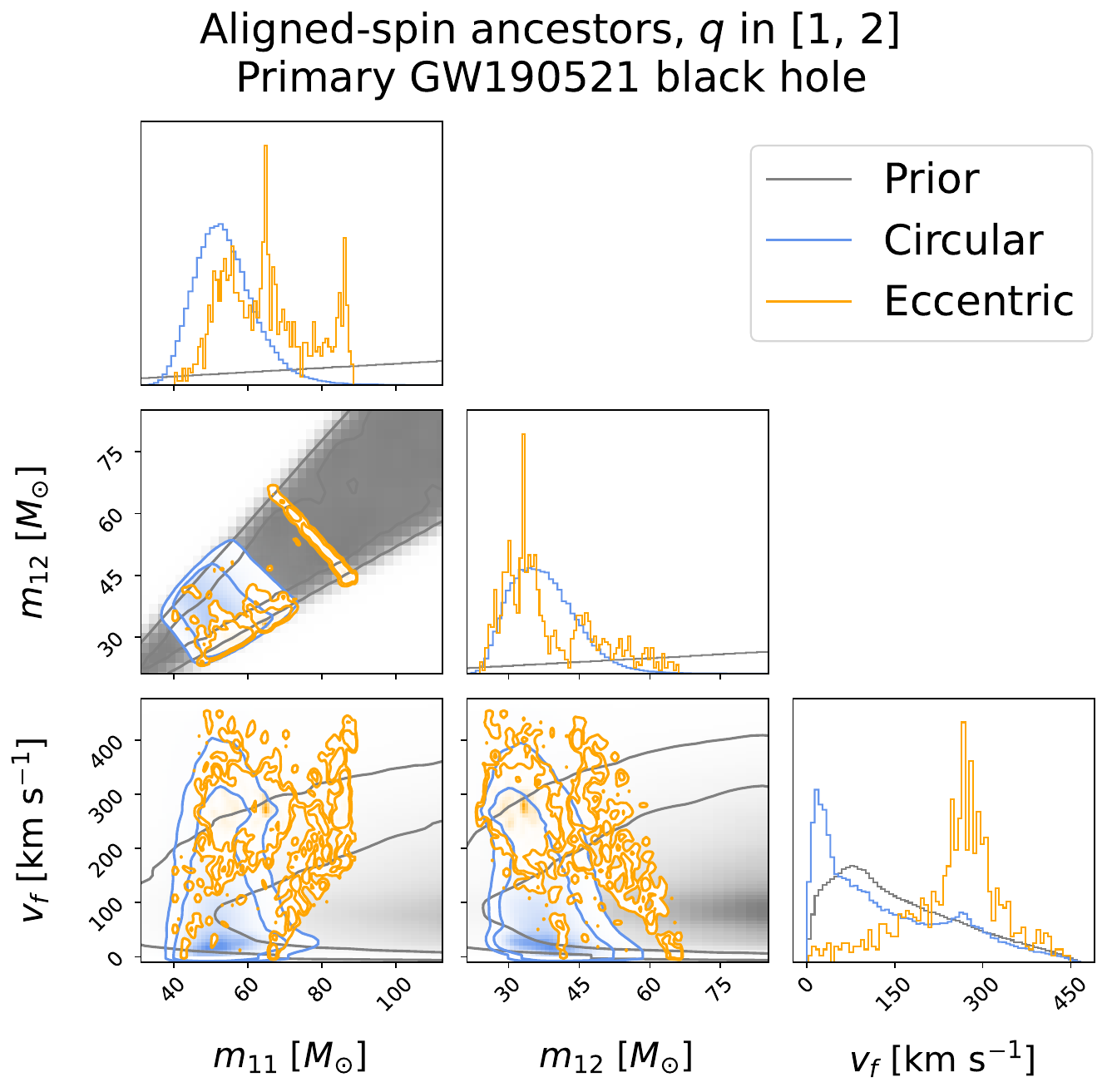}
        \includegraphics[width=0.33\textwidth]{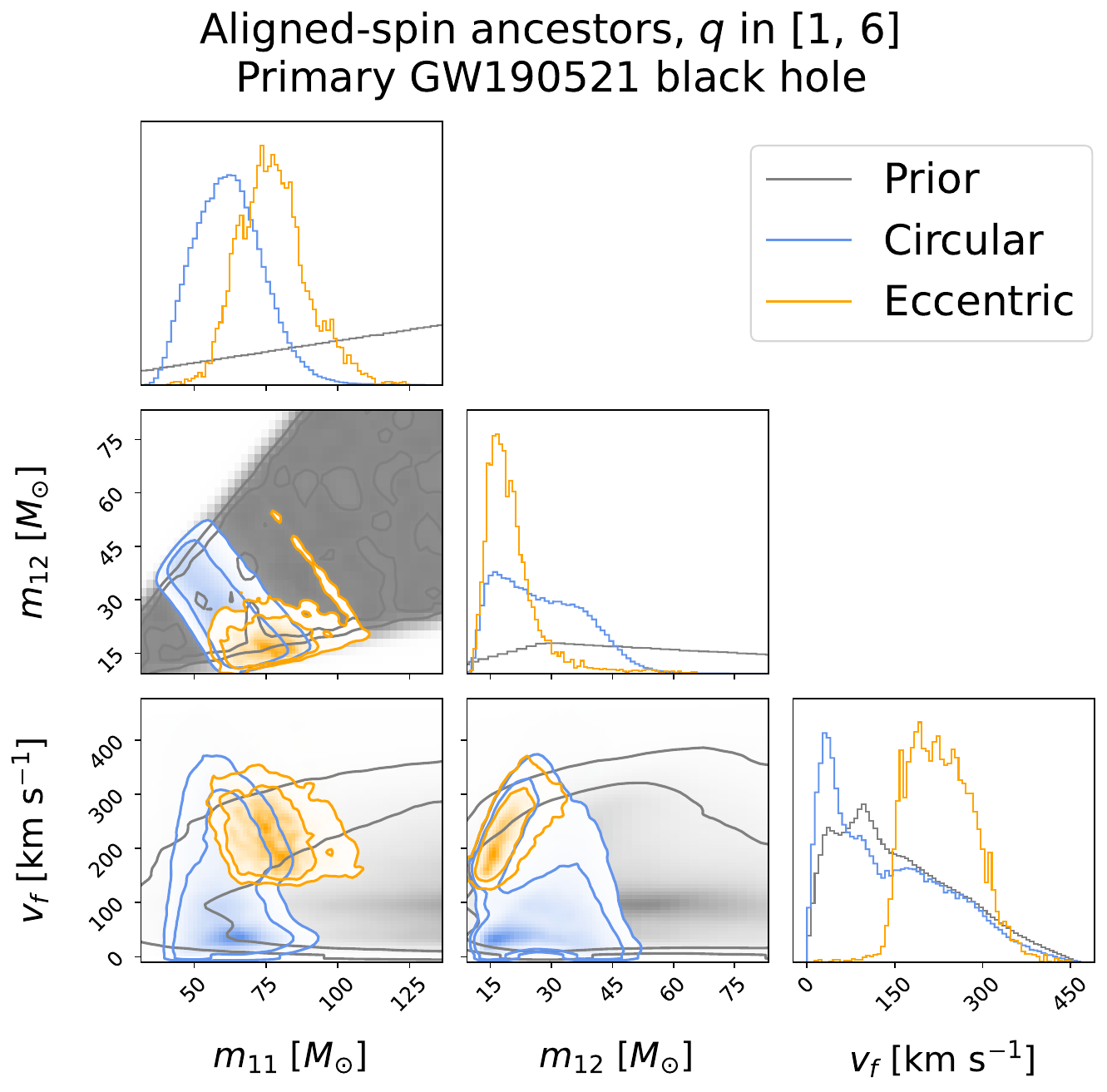}
        \caption{\textbf{Posterior distributions for the ancestral masses and birth kicks of the primary component black hole of GW190521.} We show the posterior two-dimensional $68\%$ and $90\%$ credible regions, together with the corresponding one-dimensional posterior distributions. The left panel corresponds to a \textcolor{black}{$p(\theta_{\rm anc})$} distribution allowing for precessing ancestors while the central and right \jc{panels} correspond to aligned-spin ancestors, respectively restricted to mass ratios $q\leq 2$ and $q \leq 6$. Results obtained under the original LVK analysis of GW190521 are shown in blue while those obtained under the eccentric merger analysis by Romero-Shaw et. al. are shown in orange. Prior distributions are shown in gray.}
        \label{FIG: BH1_randoms_spins}
    \end{figure*}

    \begin{figure*}
        \includegraphics[width=0.33\textwidth]{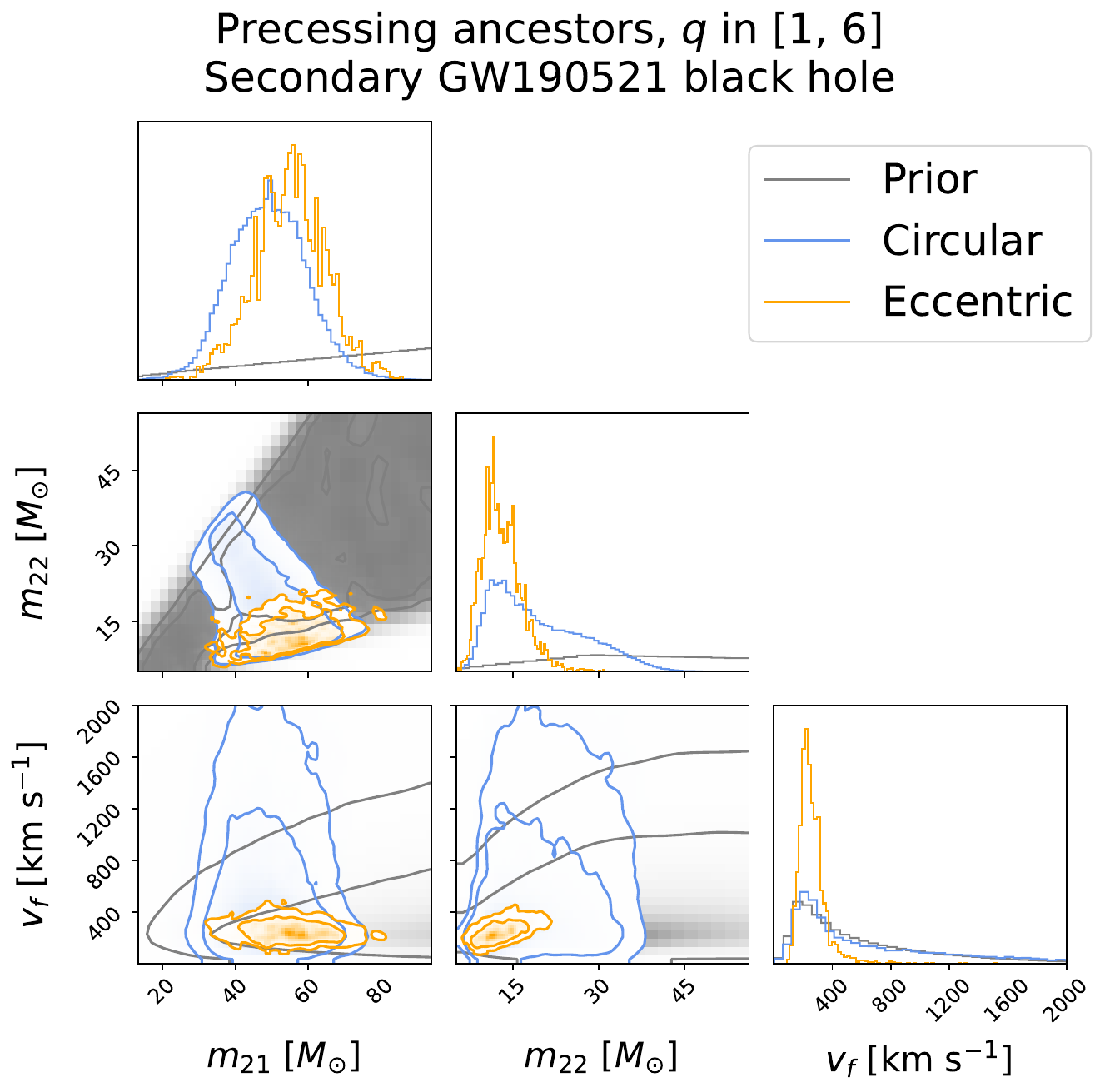}
        \includegraphics[width=0.33\textwidth]{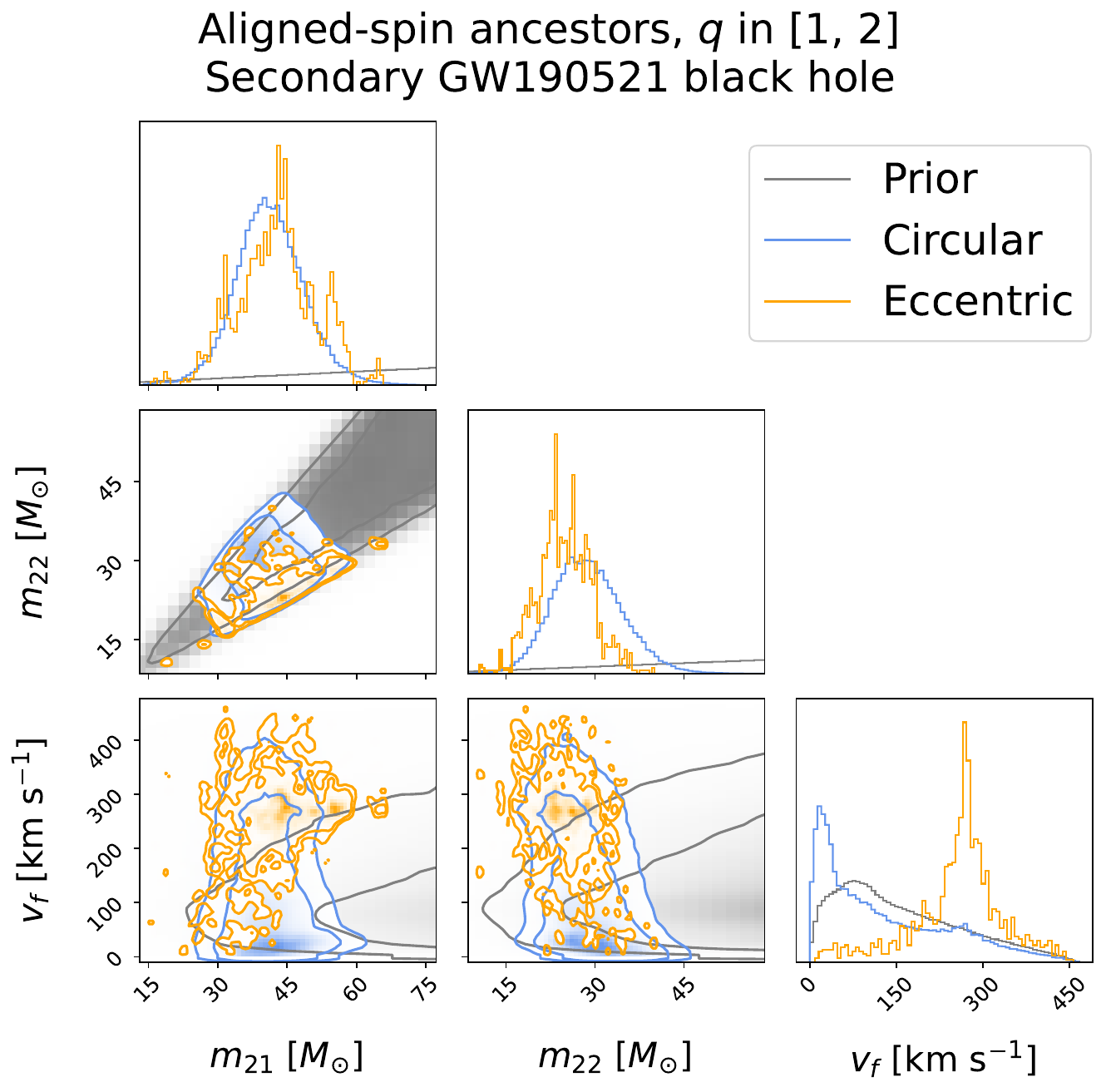}
        \includegraphics[width=0.33\textwidth]{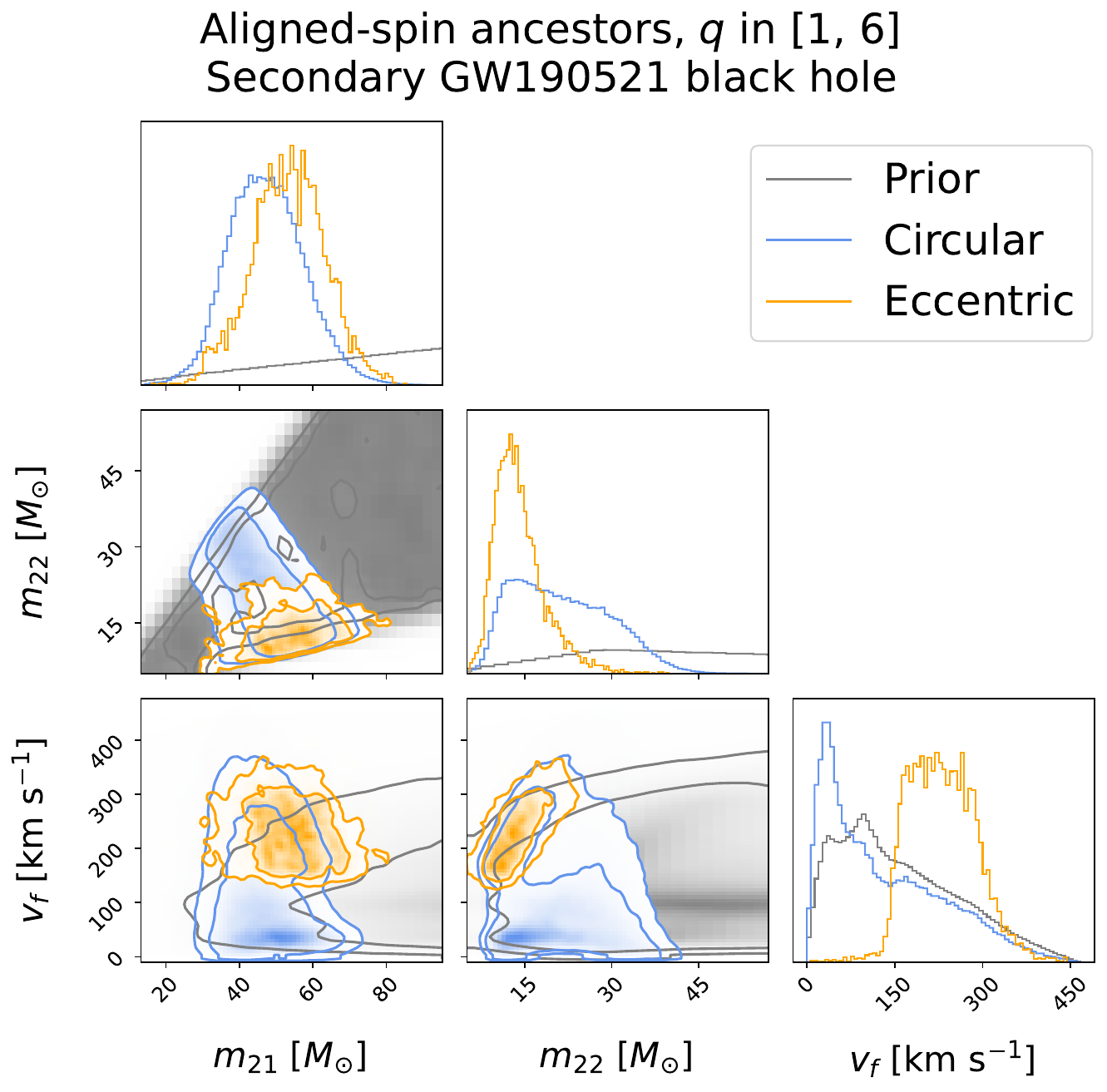}
        \caption{\textbf{Posterior distributions for the ancestral masses and birth kicks of the secondary component black hole of GW190521.} Same as Figure \ref{FIG: BH1_randoms_spins} but for the secondary component of GW190521.}
        \label{FIG: BH2_randoms_spins}
    \end{figure*}

    \subsection{Final spin vs. final kick relation}

        Figure \ref{FIG: Prior_distribution} shows the final spin and kick distributions obtained under each of our three \textcolor{black}{$p(\theta_{\rm anc})$} distributions for the ancestral BBHs. The left panel corresponds to our ``precessing \textcolor{black}{distribution}''. The central and right panels corresponds to our two ``aligned-spin'' \textcolor{black}{distributions}. There are several aspects that must be highlighted. First, a tight relation between the final spin and kick can be observed. This clearly shows that accurate estimates of the ``child'' BH spin magnitude can place strong constraints on its putative birth kick. Second, for our precessing \textcolor{black}{$p(\theta_{\rm anc})$}, low spins impose significantly tighter (lower and upper) bounds on the birth kick than large spins. In particular, note that while spins smaller than $\simeq 0.2$ lead to kicks below $500$ km/s (which would be retained in Milky Way like galaxies and denser) these must also surpass $\simeq 100$ km/s, enough to expel the final BH from any globular cluster. A qualitatively similar situation is observed for our ``aligned spin'' \textcolor{black}{$p(\theta_{\rm anc})$}, with the difference that kicks are capped at $500$ km/s due to the absence of precession.  
        
        In all the above cases, an interesting conclusion can be obtained. The observation of a BBH with a component BH populating the PISN gap and a spin \cor{$a \leq 0.3$} in a host environment with $v_{esc}< 100km/s$, like a Globular Cluster, would indicate that the PISN gap is being populated by mechanisms other than hierarchical BBH formation. We note that a similar conclusion was also obtained by \cite{Gerosa2021_heavyslow}, who found that the formation of BHs with low spins requires of very fine-tuned ancestral properties. Here, we reinforce this idea \textit{adding the kick information and} showing that, even if these are formed, the will most-likely escape the cluster. 

    \subsection{Spin magnitudes of the GW190521 component black holes} 

        We now discuss the estimates for the spin magnitudes of the component BHs of GW190521 \textcolor{black}{inferred by our input analyses}, which are key to infer the putative birth kicks. The left and right panels of Figure \ref{FIG: Spin_posterior} show respectively the posterior distributions for the spin magnitudes of the primary and secondary component BHs of GW190521. Blue posteriors correspond the original analysis performed by the LVK \cite{GW190521D}, using the state-of-the-art model for quasi-circular precessing BBHs \texttt{NRSur7dq4} \cite{NRSur7dq4}. Orange posteriors, instead, correspond to the analysis by Romero-Shaw et. al. \cite{Isobel_ecc} using a non-precessing model for BBHs with eccentricities in the range $e_{10}=[0,0.2]$ \cite{Cao2017,Liu_SEOBNRE}, estimated at an orbital frequency $f_{\rm ref}=10$ Hz. 
        
        From the above posteriors, it is rather direct to expect that birth-kick estimates will not be particularly informative under the quasi-circular scenario for GW190521. The reason is that while the posteriors for both component BHs clearly rail against the Kerr limit, these are somewhat \jc{flat}, offering good support for all spin magnitudes. In contrast, under the eccentric hypothesis, both spins are constrained to values below $\sim 0.4$, clearly peaking at $0$, which will lead to strong constraints on the possible birth kicks.

    \begin{deluxetable*}{ccccccccccc}
        \label{TABLE: parameters_table}
        \tablecolumns{10}
        \tablecaption{90\% credible intervals for the masses and spins of the ancestors and birth kick of the component black holes of GW190521 for our precessing-ancestor \textcolor{black}{distribution $p(\theta_{\rm anc})$} (PP) and our aligned-spin ones (APq2, APq6). We display also the corresponding effective-spin and effective precessing-spin. The asterisk accompanying the APq2 \textcolor{black}{distribution for the eccentric scenario for GW190521, highlights the fact that only a particularly small fraction $p(\rm anc | d)$ of the posterior samples for GW190521 admit ancestors given such choice of $p(\theta_{\rm anc})$}.}
        \tablehead{
             &  GW190521 Scenario & \textcolor{black}{$p(\theta_{anc})$} & \textcolor{black}{$p(\rm anc | d)$} & $m_{1}$ ($M_{\odot}$) & $m_{2}$ ($M_{\odot}$) & $v_f$ (km/s) & $a_{1}$ & $a_{2}$ & $\chi_{eff}$ & $\chi_{p}$ }
             \startdata 
             &          &  PP  & 0.97 & $64^{+20}_{-18}$ & $22^{+21}_{-10}$ & $404^{+1423}_{-271}$ & $0.75^{+0.24}_{-0.64}$ & $0.52^{+0.43}_{-0.47}$ & $0.10^{+0.64}_{-0.74}$ & $0.39^{+0.45}_{-0.29}$  \\ [1.5ex]
             & Circular & APq2 & 0.86 & $53^{+15}_{-11}$  & $36^{+13}_{-9}$    & $104^{+240}_{-93}$      &  $0.72^{+0.26}_{-0.64}$  &  $0.59^{+0.38}_{-0.53}$ &  $0.21^{+0.70}_{-1.06}$ & $N/A$ \\ [1.5ex]
             &          & APq6 & 1.00 & $61^{+20}_{-17}$ & $26^{+19}_{-13}$ & $117^{+208}_{-100}$      &  $0.66^{+0.32}_{-0.58}$  &  $0.52^{+0.44}_{-0.46}$ &  $0.15^{+0.71}_{-0.89}$ & $N/A$  \\ [1.5ex]
         BH$_1$  & & & & &\\ [1.5ex]
             &                          &  PP & 0.73 & $78^{+27}_{-16}$    &   $17^{+9}_{-5}$  &   $237^{+147}_{-81}$     &  $0.70^{+0.27}_{-0.44}$ &  $0.53^{+0.43}_{-0.47}$ & $-0.55^{+0.42}_{-0.23}$  &    $0.16^{+0.24}_{-0.10}$ \\ [1.5ex]
             &  Eccentric&       APq2$^{*}$ & 0.16 & $64^{+22}_{-15}$    &   $35^{+23}_{-8}$ &   $264^{+119}_{-178}$ & $0.89^{+0.10}_{-0.39}$ & $0.69^{+0.29}_{-0.60}$ & $-0.77^{+0.42}_{-0.20}$ & $N/A$  \\ [1.5ex]
             &                          &  APq6 & 1.00  &$76^{+22}_{-16}$    &   $18^{+14}_{-5}$ &   $225^{+98}_{-74}$ & $0.75^{+0.23}_{-0.41}$ & $0.50^{+0.45}_{-0.46}$ & $-0.60^{+0.35}_{-0.28}$ & $N/A$ \\ [1.5ex]
            \hline \vspace{-0.6cm}\\ [1.5ex]
             & &  PP & 0.97 &$49^{+17}_{-16}$    &   $17^{+17}_{-8}$ &   $437^{+1415}_{-307}$ & $0.77^{+0.22}_{-0.66}$ & $0.51^{+0.44}_{-0.45}$ & $0.16^{+0.59}_{-0.77}$ & $0.40^{+0.44}_{-0.30}$ \\ [1.5ex]
             &  Circular & APq2  & 0.88 &  $41^{+12}_{-11}$ & $28^{+10}_{-8}$ & $96^{+244}_{-87}$ & $0.73^{+0.25}_{-0.65}$ & $0.60^{+0.37}_{-0.54}$ & $0.27^{+0.65}_{-1.10}$ & $N/A$ \\ [1.5ex]
             & & APq6  & 1.00 &   $47^{+17}_{-15}$ & $20^{+15}_{-10}$ & $104^{+218}_{-88}$ & $0.67^{+0.31}_{-0.59}$ & $0.52^{+0.44}_{-0.47}$ & $0.23^{+0.64}_{-0.95}$ & $N/A$ \\ [1.5ex]
              BH$_{2}$  & & & & \\ [1.5ex]
             & &  PP & 0.66 &$55^{+15}_{-16}$    &   $12^{+6}_{-4}$  &   $243^{+175}_{-89}$ & $0.70^{+0.27}_{-0.47}$ & $0.49^{+0.45}_{-0.43}$ & $-0.53^{+0.42}_{-0.26}$ & $0.17^{+0.27}_{-0.10}$\\ [1.5ex]
             &  Eccentric &APq2$^{*}$  & 0.20 & $42^{+13}_{-13}$    &   $24^{+7}_{-7}$ &   $267^{+115}_{-179}$ & $0.92^{+0.08}_{-0.39}$ & $0.71^{+0.27}_{-0.61}$ & $-0.81^{+0.45}_{-0.16}$ & $N/A$ \\[1.5ex]
             & & APq6  &  1.00 &  $53^{+15}_{-17}$    &   $13^{+9}_{-4}$ &   $228^{+98}_{-77}$ & $0.75^{+0.23}_{-0.44}$ & $0.50^{+0.45}_{-0.45}$ & $-0.59^{+0.37}_{-0.28}$ & $N/A$ \\ [1.5ex]
            \enddata
    \end{deluxetable*}

    \begin{deluxetable*}{cccccccc}
        \label{TABLE: probability_table}
        \tablecolumns{10}
        \tablecaption{Probability \textcolor{black}{$p(2g|\rm anc)$} for the ancestor components residing below the PISN mass gap at the same time that birth kick is lower that the typical escape velocity of a Globular Cluster (${\rm GC}$), the Milky Way (${\rm MW}$), a Nuclear Star Cluster (${\rm NSC}$) or an Elliptical Galaxy (${EG}$). 
        \textcolor{black}{The corresponding probability that GW190521 is of second generation $p_{2g}$ is obtained multiplying $p(2g|\rm anc)$ by the value $p(\rm anc | d)$, which denotes probability that the corresponding component of GW190521 that admit ancestors within the ancestral distributions $p(\theta_{\rm anc})$, according to our input analyses, or ``scenarios''.}
        We show values obtained for two orbital scenarios for GW190521 and three choices of the  $p(\theta_{\rm anc})$ distribution. \textcolor{black}{The asterisk accompanying the APq2 distribution for the eccentric scenario for GW190521, denotes that a small fraction of the posterior samples for GW190521 $p(\rm anc | d)$ admit ancestors.}}
        \tablehead{
             & \colhead{GW190521 Scenario} & \colhead{$p(\theta_{\rm anc})$} & \colhead{$p(\rm anc | d)$} &\colhead{$p^{\rm GC}(2g|\rm anc)$} & \colhead{$p^{\rm MW}(2g|\rm anc)$} & \colhead{$p^{\rm NSC}(2g|\rm anc)$} & $p^{\rm EG}(2g|\rm anc)$}
             \startdata
             &                          & PP   & 0.97 & 0.002 & 0.293 & 0.461 & 0.518  \\ 
             & Circular                 & APq2 & 0.86& 0.259 & 0.903 & 0.903 & 0.903  \\
             &                          & APq6 & 1.00 & 0.135 & 0.605 & 0.605 & 0.605  \\
         BH$_{1}$ & & & & & & \\
             &                          & PP   & 0.73 & 0.000 & 0.114 & 0.116 & 0.116 \\
             &  Eccentric               & APq2* & 0.16 & 0.011 & 0.572 & 0.572 & 0.572 \\
             &                          & APq6 & 1.00 & 0.000 & 0.143 & 0.143 & 0.143 \\
            \hline \vspace{-0.3cm} \\
             &                          & PP   & 0.97 & 0.004 & 0.563 & 0.834 & 0.927 \\ 
             &  Circular                & APq2 & 0.88 & 0.326 & 0.997 & 0.997 & 0.997 \\
             &                          & APq6 & 1.00 & 0.255 & 0.951 & 0.951 & 0.951 \\
            BH$_{2}$  & & & & \\
             &                          &  PP  & 0.66 & 0.000 & 0.839 & 0.849 & 0.849 \\
             &  Eccentric               & APq2* & 0.20 & 0.026 & 0.998 & 0.998 & 0.998 \\
             &                          & APq6 & 1.00 & 0.002 & 0.878 & 0.878 & 0.878 \\
        \enddata
    \end{deluxetable*}

    \subsection{Ancestral BBHs of GW190521}

\begin{figure}
            \includegraphics[width=0.23\textwidth]{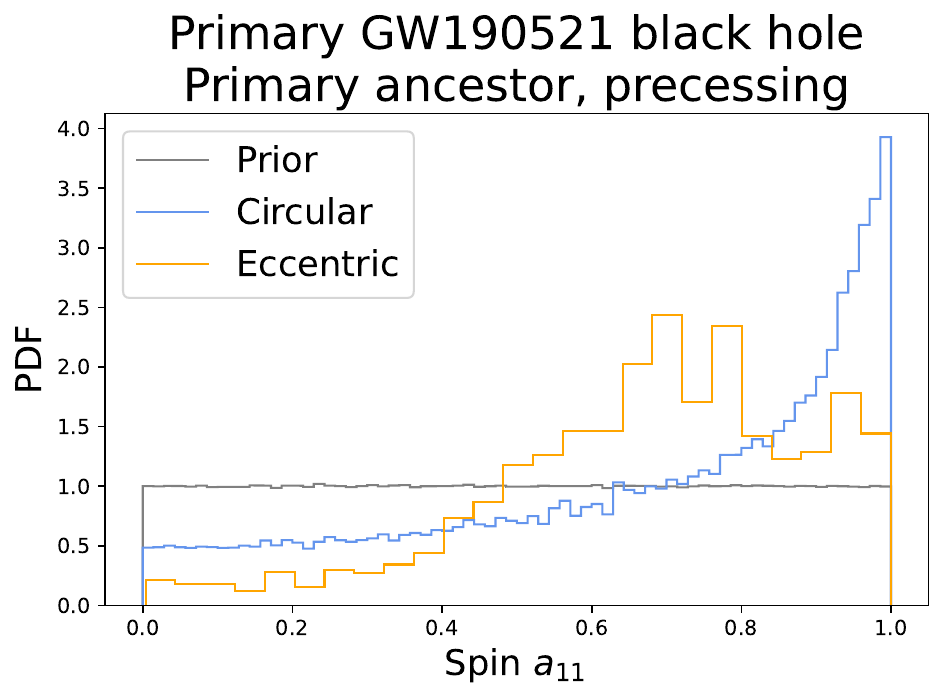}
            \includegraphics[width=0.23\textwidth]{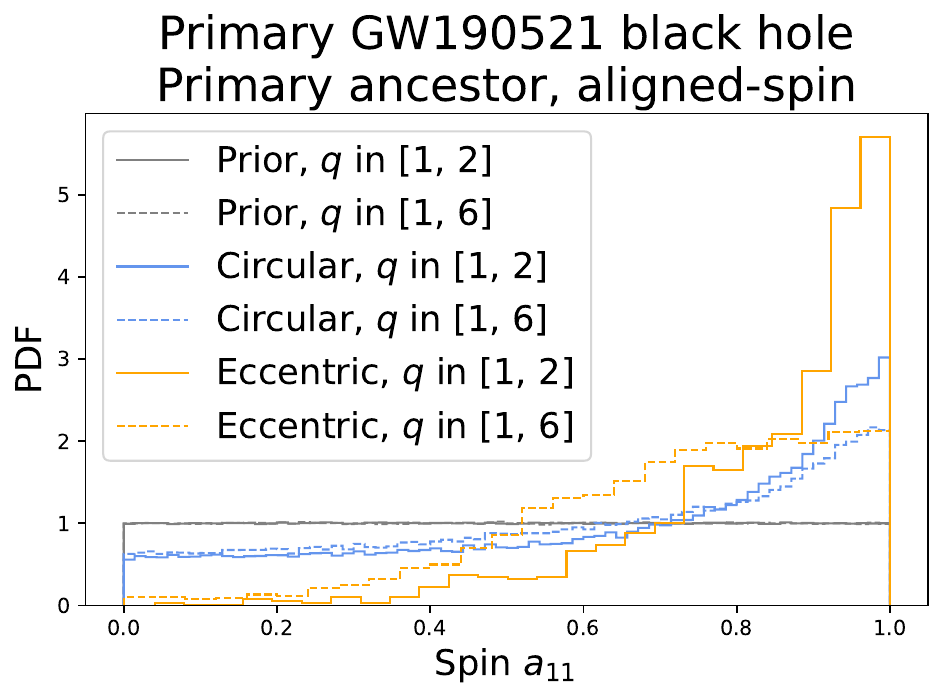}
            \includegraphics[width=0.23\textwidth]{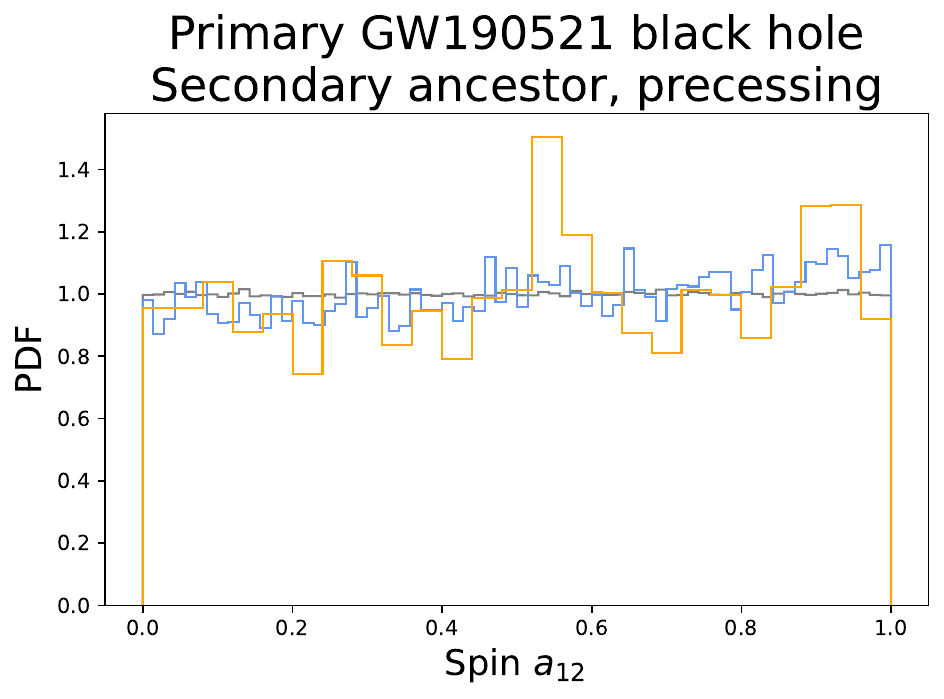}
            \includegraphics[width=0.23\textwidth]{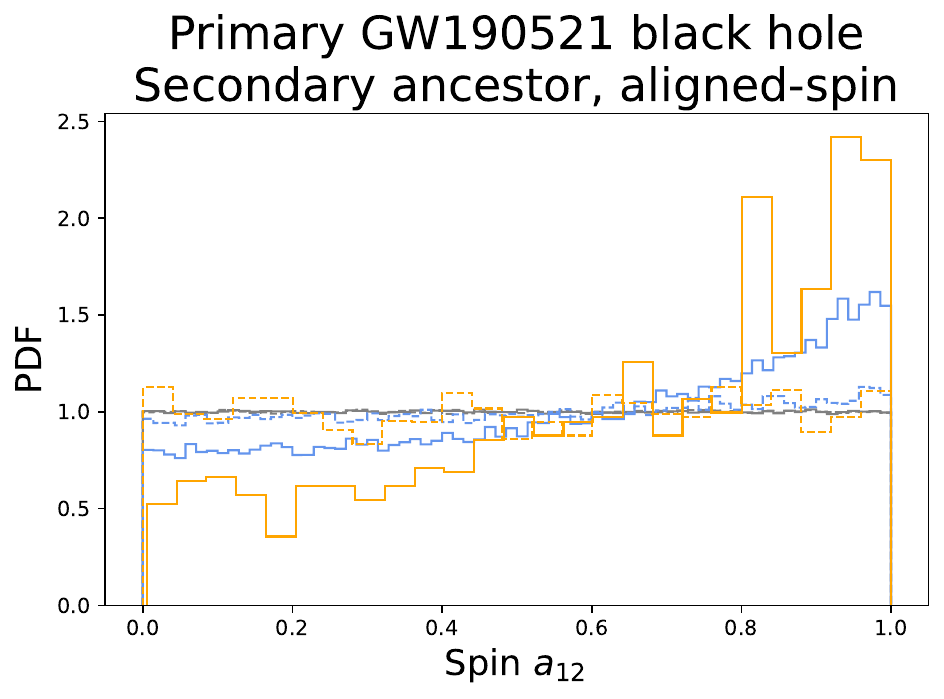}
            \includegraphics[width=0.23\textwidth]{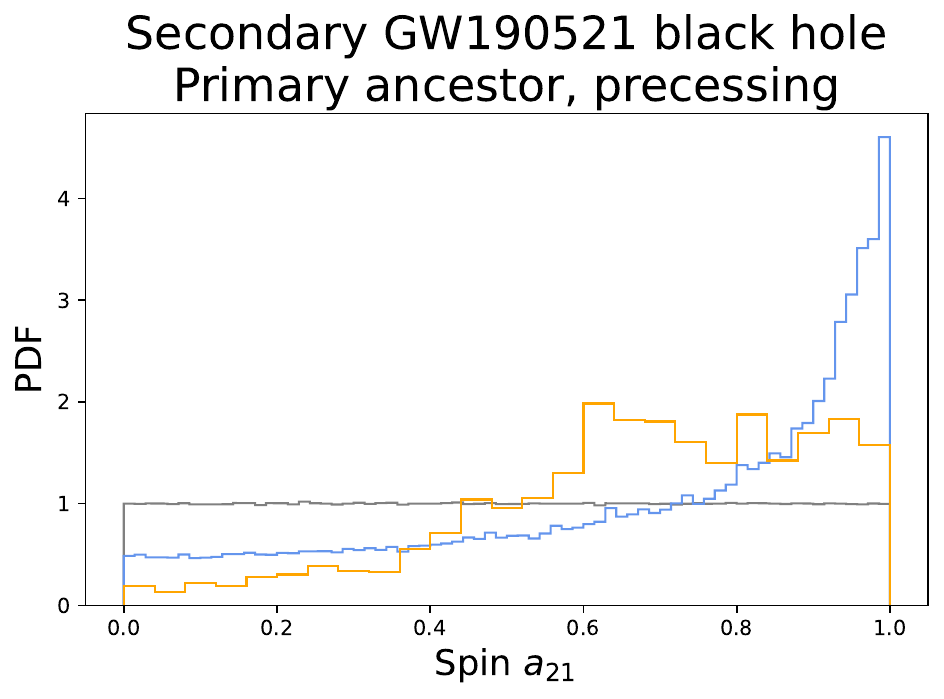}
            \includegraphics[width=0.23\textwidth]{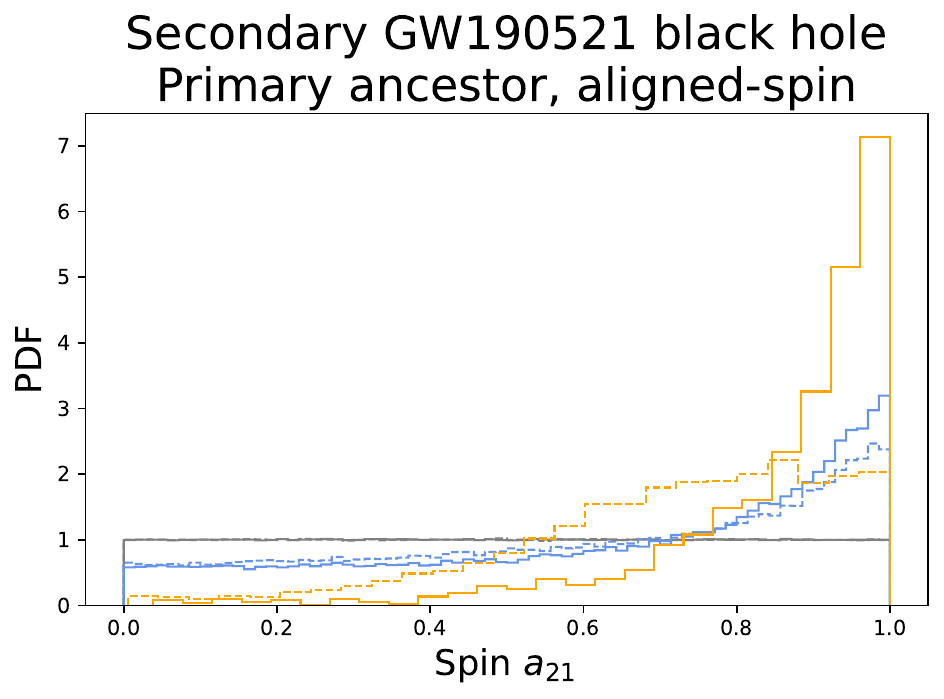}
            \includegraphics[width=0.23\textwidth]{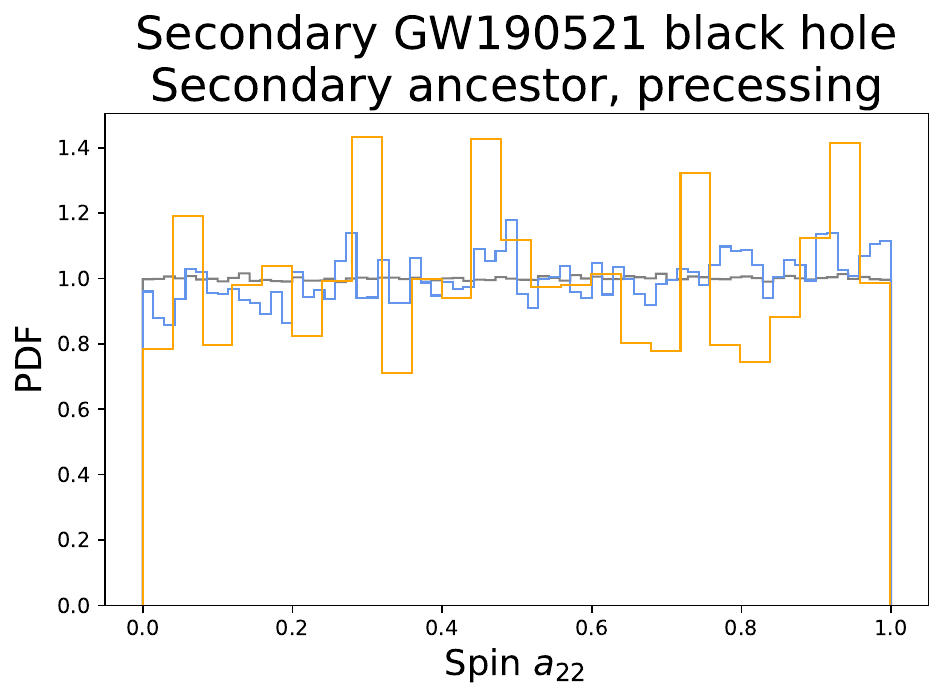}
            \includegraphics[width=0.23\textwidth]{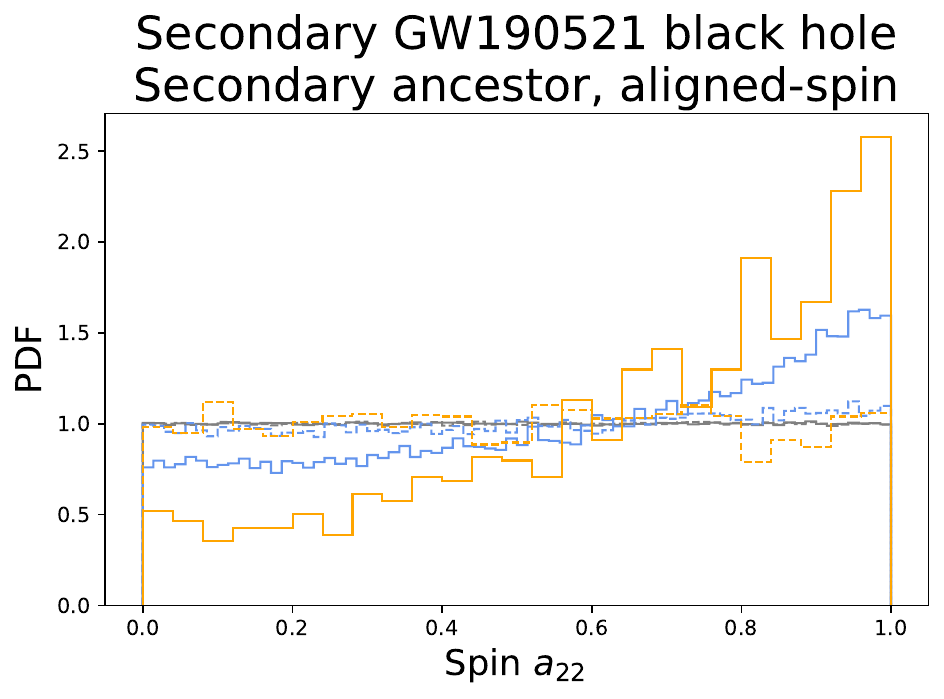}
            \caption{\textbf{Ancestral spin magnitudes.} Posterior distributions for the spin magnitudes of the ancestral BBHs of GW190521. Left panels correspond to our precessing distribution $\textcolor{black}{p(\theta_{\rm anc})}$ for the ancestors while right panels correspond to our aligned-spin ones. The blue and orange distributions correspond respectively to a quasi-circular and eccentric scenario for GW190521. Prior distributions are represented in grey.}
            \label{FIG: BH1_BH2_parental_spin}
        \end{figure}

\begin{figure*}
            \includegraphics[width=0.5\textwidth]{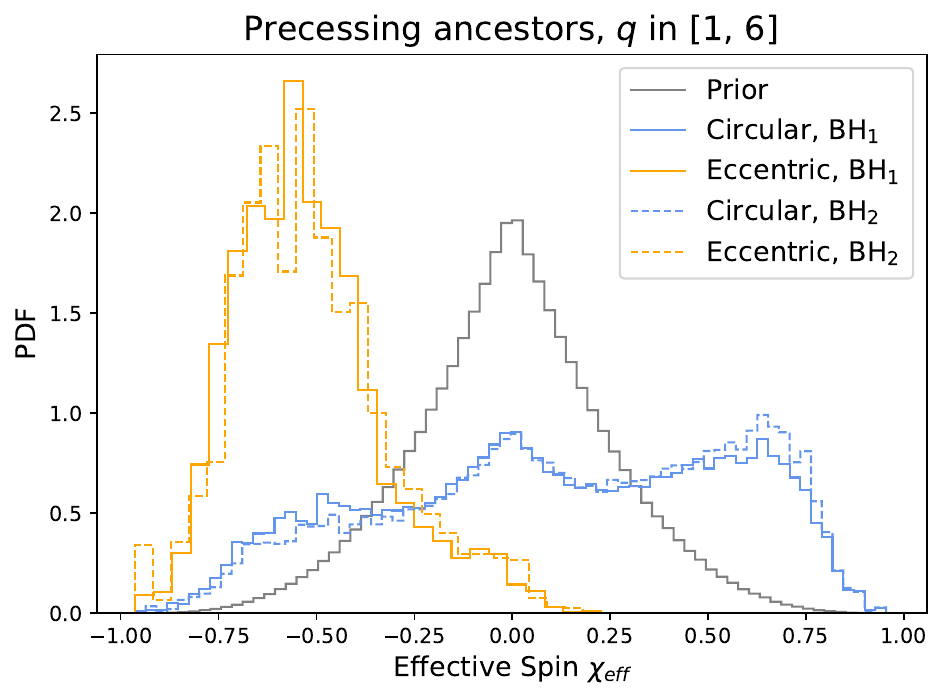}
            \includegraphics[width=0.5\textwidth]{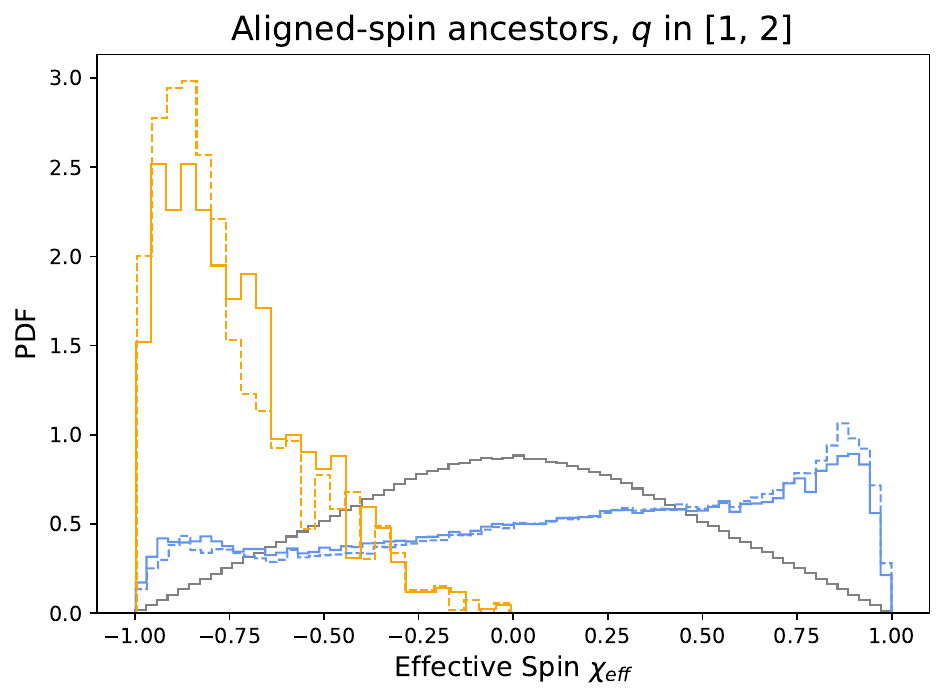}
            \includegraphics[width=0.5\textwidth]{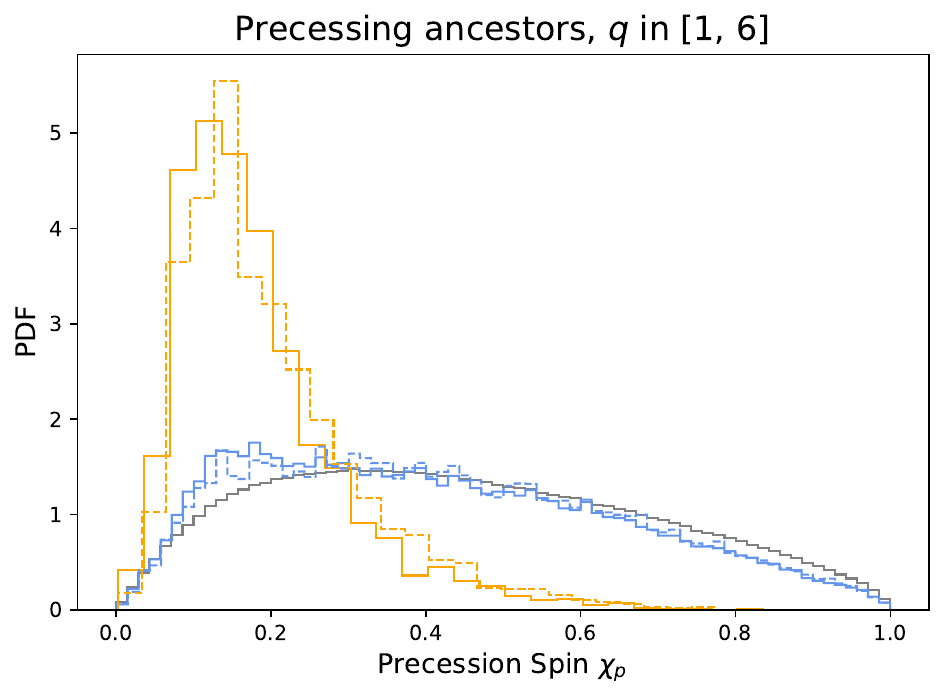}
            \includegraphics[width=0.5\textwidth]{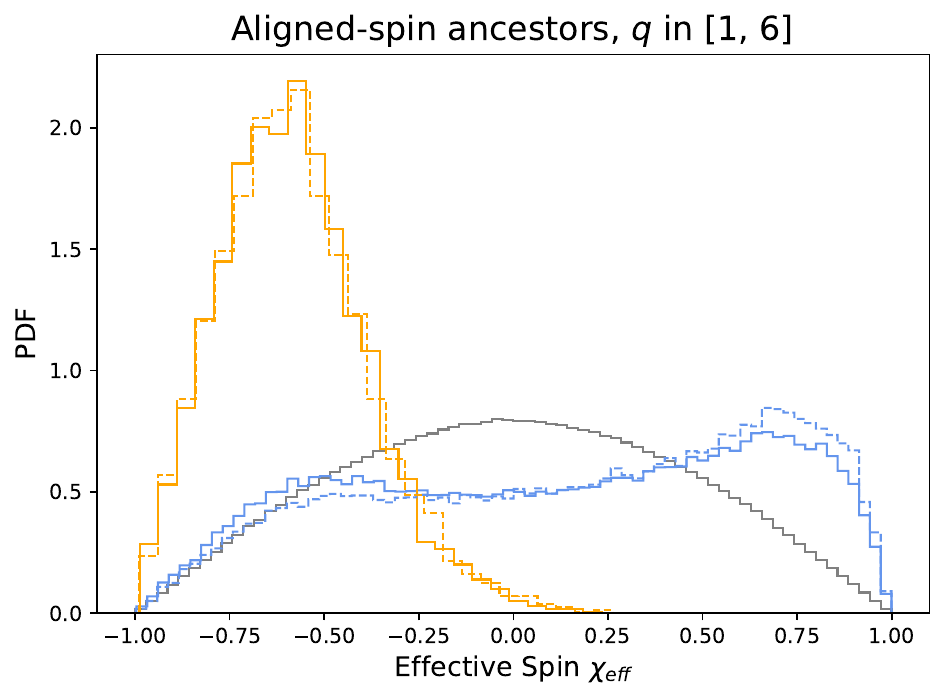}
            \caption{\textbf{Ancestral effective spin parameters.} Same as Figure \ref{FIG: BH1_BH2_parental_spin} for the effective spin $\chi_{\rm eff}$ and effective spin-precession $\chi_p$ parameters of the ancestral black-hole binaries of GW190521.}
            \label{FIG: BH1_BH2_parental_spin_eff}
        \end{figure*}

        Figure \ref{FIG: BH1_randoms_spins} shows the posterior two-dimensional $90\%$ credible regions, together with the corresponding one-dimensional posterior distributions, for the ancestral masses and birth kick of the primary BH of GW190521. Figure \ref{FIG: BH2_randoms_spins} shows the same for the secondary BH. The left panel shows results obtained under our ``precessing'' \textcolor{black}{distribution} for the ancestral black holes while the central and right panels show those corresponding to our ``aligned-spin'' ones, respectively restricted to mass ratios $q\leq 2$ and $q\leq 6$. The results in blue and orange respectively correspond to the hypothesis that GW190521 was a quasi-circular and an eccentric BBH \footnote{\jc{The orange posteriors are notably less smooth due to the fact that these were obtained through a likelihood re-weighting method \cite{Payne2019_rew}, after an initial sampling making use of a non-eccentric waveform model, which can lead to an important sample reduction.}}. \textcolor{black}{Our ancestral $p(\theta_{\rm anc})$} distributions are represented in grey. We note that we have restricted the plot ranges to values encompassing our posteriors, to improve the visibility of the results.  Next, Fig. \ref{FIG: BH1_BH2_parental_spin} shows the posterior distributions for the individual spin magnitudes of the ancestral BHs. The left (right) panels correspond to our precessing (aligned-spin) ancestral \textcolor{black}{distributions $p(\theta_{\rm anc})$}. Finally, Fig. \ref{FIG: BH1_BH2_parental_spin_eff} shows the posterior distributions for the effective-spin parameter $\chi_{\rm eff}$ \cite{Santamaria:2010yb} and the effective precessing-spin parameter $\chi_p$ \cite{Schmidt:2012rh,Schmidt:2014iyl}. All results are reported numerically in Table \ref{TABLE: parameters_table}, in terms of median values accompanied by symmetric $90\%$ credible intervals. As expected, our results show a strong dependence on both the choice of \textcolor{black}{$p(\theta_{\rm anc})$} and the scenario considered for GW190521, as we will discuss in detail. While we report results for the putative ancestors of both components of GW190521, in the following we will mostly focus our discussion on those of the primary BH, as this has an extremely low probability of avoiding the PISN gap. 

    \subsubsection{LVK quasi-circular precessing scenario}

        Under our precessing \textcolor{black}{$p(\theta_{\rm anc})$ distribution}, we find ancestral masses for the primary BH $(m_{11},m_{12})=(64^{+20}_{-18}, 22^{+21}_{-10})M_\odot$. \jc{Given our reference range for the PISN gap, we obtain a $\sim 50\%$ chance that both ancestors reside below it}. For the primary ancestor, we obtain an informative posterior distribution for the spin magnitude $a_{11}$ that rails against the Kerr limit, similar to that for $a_1$ itself. In contrast, we obtain \jc{a} totally uninformative posterior for $a_{12}$, owing to its smaller impact on the resulting final spin. As per the spin orientations, we obtain a quite unconstrained but informative posterior for $\chi_{\rm eff}$. This deviates from the distribution induced by $p(\theta_{\rm anc})$ through peaks at highly positive and negative $\chi_{\rm eff}$, while showing less support than $p(\theta_{\rm anc})$ for nearly null $\chi_{\rm eff}$. This result reflects the fact that the production of nearly null (extremal) remnant spins, both well supported by the $a_1$ posterior, requires highly negative (positive) ancestral $\chi_{\rm eff}$. In contrast, the distribution \textcolor{black}{$p(\theta_{\rm anc})$} (which peaks at $\chi_{\rm eff} =0$) highly suppresses such remnant spin values, showing a preference for $a_f \in [0.4,0.6]$ (see Figure \ref{FIG: Prior_distribution}). Finally, we obtain a completely uninformative posterior for $\chi_p$. The latter, together with the uninformative posterior for $a_1$, leads to a highly unconstrained posterior for the putative birth recoil $v_f=404^{+1423}_{-271}$km/s. This yields almost zero chances for \jc{this} BH to be retained in a Globular Cluster, which are typically considered to have maximal escape velocities of $50$ km/s \cite{Baumgardt2018}. While we defer detailed explanations \jc{to} the next section, the latter already shows the dramatic role of the recoil velocity in this type of \jc{studies}.\\
        
        Our aligned-spin \textcolor{black}{ancestral distribution} restricted to $q\leq2$ (APq2) leads to slightly smaller(larger) ancestral primary(secondary) masses $(m_{11},m_{12})=(53^{+15}_{-11},36^{+13}_{-9})M_\odot$, notably raising the probability that both ancestral masses lay below the PISN gap. This is simply due to the limitation imposed in the ancestral mass ratio. We note, however, that \textcolor{black}{under this ancestral distribution, black holes with spin magnitudes $a_f \leq 0.2$, which represent $14\%$ of the posterior distribution for the primary BH of GW190521, do not admit ancestors, yielding $p(\rm anc | d) = 0.86$}. Expanding the allowed ancestral mass ratio to $q \leq 6$, we obtain masses totally consistent with those obtained under our precessing $p(\theta_{\rm anc})$. The kick estimation, however, is radically different, owing to the smaller kicks allowed by aligned-spin mergers. In particular, we obtain $v_f=104^{+240}_{-93}$ km/s and $v_f=117^{+208}_{-100}$ km/s under each of our aligned-spin priors. While, once again, this essentially \jc{ensures} the ejection of the final BH from a Globular Cluster, the probability of retention in denser environments is significantly increased, as we will later discuss. \\

        For the ancestors of the secondary BH of GW190521, we obtain similar results of which we highlight two aspects. First, while the mass estimates are \jc{naturally} smaller, we still obtain a non-zero chance for the primary ancestor to be within the PISN gap. Second, we also obtain non-zero chances for the recoil to be below the $50$ km/s limit characteristic of globular clusters.
     
    \subsubsection{Eccentric non-precessing scenario}

        The assumption of this scenario for GW190521 has dramatic consequences due to the significantly different spin estimates it leads to. First, we note that under our APq2 aligned-spin \textcolor{black}{$p(\theta_{\rm anc})$ more than $80\%$ of the posterior distribution for both GW190521 components do not admit an ancestor, due to incompatibility of their spin magnitudes with those arising from the remnants of the ancestral distribution. This sets a limit of $p(\rm anc | d) < 0.2$ on the probability $p_{2g}$ that the GW190521 components are of second generation. Under our precessing $p(\theta_{\rm anc})$ -- for which $p(\rm anc | d) = 0.73$ --} we obtain ancestral masses for the primary component of GW190521 $(m_{11},m_{12})=(78^{+27}_{-16}, 17^{+9}_{-5})M_\odot$. Importantly, this leaves only a $\simeq 10\%$ chance for both ancestral masses to reside below the PISN gap, much lower than the $\simeq 50\%$ chance obtained under the LVK quasi-circular scenario. Equivalent results are obtained for the masses under the aligned-spin APq6 ancestral distribution. For the ancestral spins, as expected looking at the $a_1$ posterior, we obtain results that are very different from those of the quasi-circular case. First, we obtain an informative estimate $a_{11}=0.70^{+0.27}_{-0.44}$, clearly peaking around $a_{11}=0.7$, together with an uninformative distribution for $a_{12}$. Second, we obtain informative posteriors for $\chi_{\rm eff}$ and $\chi_{\rm p}$. These respectively yield $\chi_{\rm eff}=-0.55^{+0.42}_{-0.23}$ -- essentially ruling out positive values -- and $\chi_{p}=0.16^{+0.24}_{-0.10}$, which rather discards strong precession. As discussed earlier, these informative posteriors arise from the informative posterior for $a_1$, which peaks at $a_1=0$ and \jc{is negligible beyond} $a_1=0.3$. On the one hand, the production of such remnant BHs, \jc{requires} highly negative effective spins. On the other hand, the latter also requires the spins to form angles of nearly 180 degrees with the orbital angular momentum (i.e., the spins must be essentially anti-aligned with the orbital angular momentum), which prevents the system from having large in-plane components and, therefore, displaying strong precession. These properties lead to a \textit{highly informative kick estimate} $v_f=237^{+147}_{-81}$km/s, dramatically different from that inferred under the LVK quasi-circular scenario. This means that the upper bound for the birth kick is significantly decreased with respect to the LVK case, therefore increasing the chances of retention in dense environments, even under the precessing ancestral prior. On the contrary, we find that assuming an aligned-spin ancestral prior, the lower bound for the birth kick $225^{+98}_{-74}$ km/s is much larger \textcolor{black}{than} that obtained under LVK quasi-circular scenario for GW190521 (owing to the $\chi_{\rm eff}$) posterior, making it impossible to retain the final BH in a Globular Cluster.\\

        Regarding the putative ancestors of the secondary BH, we find again a non-zero chance for the primary ancestor to reside in the PISN gap. However, just as for the primary BH, we estimate a birth kick of $v_f={243^{+175}_{-89}}$ km/s that cannot be retained in a Globular Cluster.  
    
    \subsection{Viability of the GW190521 components as second-generation black holes}

        We now evaluate the probability $p_{2g}$ that the component BHs of GW190521 can be the remnant of a previous merger of stellar-origin BHs. To \jc{do} this, \textcolor{black}{we evaluate eq. (12) for each component BH and each ancestral distribution $p(\theta_{\rm anc})$, as function of the escape velocity of various types of environments.}\\ 
    
\begin{figure*}
            \includegraphics[width=0.5\textwidth]{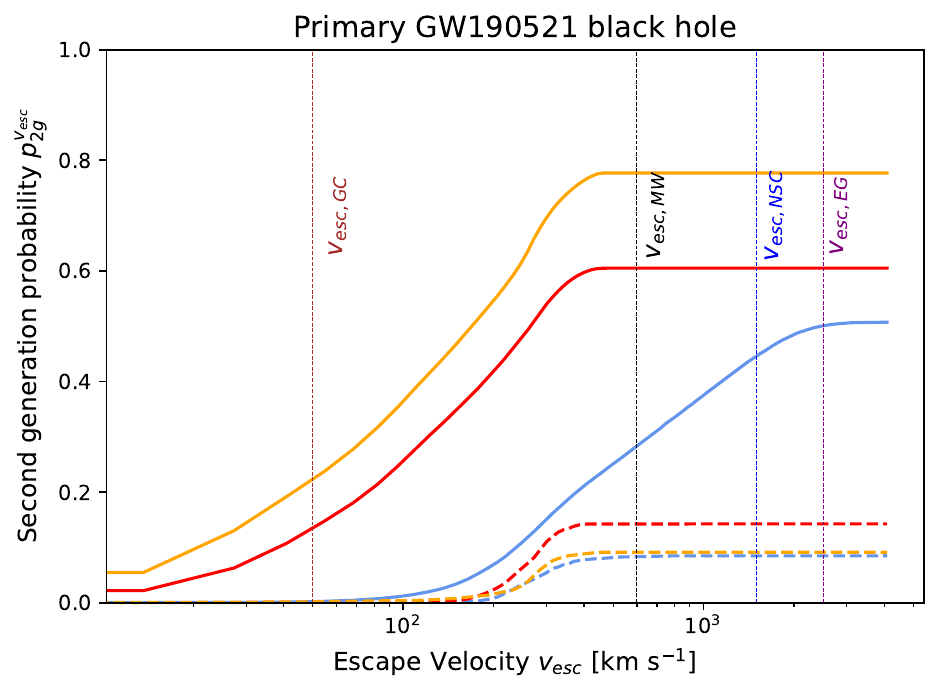}
            \includegraphics[width=0.5\textwidth]{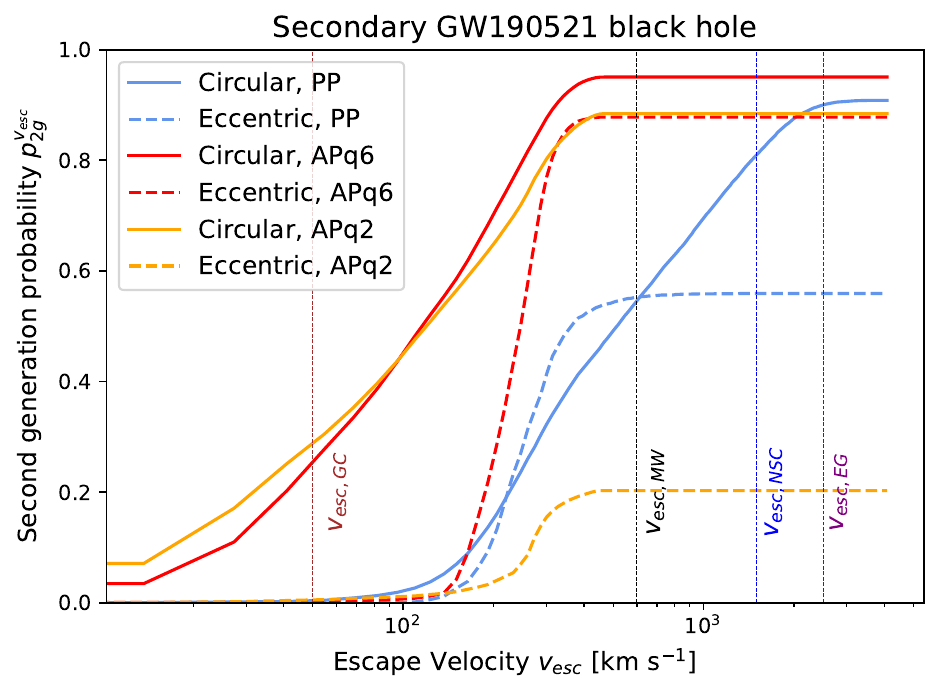}
            \caption{\textbf{Viability of the components of GW190521 as the result of a merger of stellar-origin black holes as a function of the escape velocity of the host environment}. The left (right) panel corresponds to the primary (secondary) black hole. Solid (dashed) curves correspond to a quasi-circular (eccentric) interpretation for GW190521. Blue (red) curves correspond to our precessing (aligned-spin) \textcolor{black}{$p(\theta_{\rm anc})$ distribution} for the ancestor black holes. We highlight typical maximal values for four different types of host environments, namely Globular Clusters, the Milky Way, Nuclear Star Clusters and Elliptical Galaxies.}
            \label{FIG: BH1_BH2_retention_probability}
        \end{figure*}

        Figure \ref{FIG: BH1_BH2_retention_probability} shows $p_{2g}$ as a function of the escape velocity $v_{esc}$ for the primary (left panel) and secondary (right panel) components of GW190521. We highlight in vertical bars the maximal \jc{escape} velocities for some particular types of environments, namely Globular Clusters, Nuclear Star Clusters, Elliptical Galaxies and the Milky Way. The corresponding \textcolor{black}{numerical values of $p_{2g}$ can be read from Table \ref{TABLE: probability_table}, by simply multiplying the values of the columns $p(\rm anc | d)$ and $p(\rm 2g | anc)$}. As expected from the previous section, results vary significantly depending on the chosen combination of scenario for GW190521 and \textcolor{black}{$p(\theta_{\rm anc})$ distribution} for the ancestral BHs. Also, in order to highlight the dramatic impact of the inclusion of the recoil in the assessment of the viability of the GW190521 components as second-generation BHs, we note that the rightmost value in the panels (i.e., for $v_{\rm esc}$ equal to the maximum kick velocity that \texttt{NRSur7dq4} can predict) corresponds to the case where the kick estimate is ignored so that only the values of the masses are considered. Once again, we will mostly focus our discussion on the primary BH.

    \subsubsection{``Ignoring" the birth kick: the infinite escape velocity limit}

        For high enough escape velocities, the remnant BH is always retained within its environment. This way, \textcolor{black}{once $p(\rm anc |d)$} is computed, $p_{2g}$ is simply determined by the condition $p(m_{i1},m_{i,2})<65M_\odot$. Even in this case, the choice of scenario and $p(\theta_{\rm anc})$ dramatically impacts  $p_{2g}$. For the quasi-circular scenario, we find that $p_{2g}$ is respectively bounded by approximately 0.5, 0.6 and \textcolor{black}{0.8} under our PP, APq6 and APq2 distributions. The latter distribution yields a larger $p_{2g}$ due to the lower ancestral masses it leads to. These probabilities are all reduced to \textcolor{black}{$\simeq 0.1$} if we assume an eccentric scenario for GW190521. \footnote{We note again that in the eccentric case, and \textcolor{black}{under the APq2 ancestral distribution, $84\%$ of the posterior for the primary BH of GW190521 would not admit ancestors, yielding $p(\rm anc | d)=0.16$}}. Therefore, we conclude that if GW190521 was eccentric, the probability that the primary BH is of second-generation is bounded by $p_{2g} \leq 0.1$.  
        
    \subsubsection{Finite escape velocities: the impact of the birth kick}

        The inclusion of the birth kick strongly suppresses the viability of certain environments as hosts of GW190521 (assuming that the primary BH is of second generation) as these decrease their escape velocity. For the case of GW190521 being quasi-circular, the results are mostly driven by the choice of the ancestral \textcolor{black}{distribution}, as this determines almost completely birth-kick posterior. This is reflected, for instance, by the fact that the solid-blue curve slowly decreases for $v_{esc}<3000$ km/s -- the maximal kick we can obtain for precessing mergers -- while the orange and red ones start dropping at $v_{esc}< 400$ km/s, around the maximal kick attainable for aligned-spins. In contrast, the results under the eccentric scenario (dashed) are strongly driven by the actual data. In particular, by the strong constrain that the GW190521 spin magnitudes impose on the putative birth kicks, which get restricted to $v_f \in \sim [150,350]$ km/s even if the ancestors are allowed to precess.\\ 
    
        \textcolor{black}{For a Globular Cluster, we obtain negligible values for $p_{2g} \sim 10^{-3}$ in all cases except if GW190521 was quasi-circular and its ancestors had aligned spins. We have also checked that restricting the spin magnitudes to small or null values also leads to small values of $p_{2g} \sim 10^{-2}$ (see later). For aligned-spin ancestors, we obtain probabilities of $p_{2g} = 0.22 (0.14)$} when the ancestral mass ratio is respectively restricted to $q\leq 2 (6)$. Therefore, \textcolor{black}{\textit{given the properties of GW190521 inferred by the LVK \textcolor{black}{(under their priors)}, and if it happened in a Globular Cluster}, its primary BH can only be of second generation \textcolor{black}{with $p_{2g} > 0.1$} if its ancestors had aligned spins}. \cor{We note, however, that aligned-spins are characteristic of binaries formed in isolation -- which are not expected to lead to repeated mergers. Hierarchical formation channels in Globular Clusters, in contrast, is thought to involve mergers with randomly oriented spins.}\\ For the remaining environments, we obtain the same $p_{2g}$ as when we assume an infinite $v_{esc}$, except for the case of a quasi-circular GW190521 with precessing ancestors. For the latter, we respectively obtain \textcolor{black}{$p_{2g} = \{0.29, 0.44, 0.50\}$} for the case of the Milky Way, a Nuclear Star Cluster and an Elliptic Galaxy. \cor{Among  these environments, however, we note that only Nuclear Star Clusters is expected to be densely enough populated to enable hierarchical mergers.}\\
  
\subsubsection{Restricting $p(\theta_{\rm anc})$: small/null ancestral spins \\ and reduced mass ratio} 
        
        \textcolor{black}{Given its physical interest, we also discuss the result of imposing, in precessing ancestral distribution, restrictions on the maximal spin magnitude and mass ratio that better represent theoretically expected characteristic of stellar-born black holes. First, we restrict our spin magnitudes to both null and $a_i < 0.2$. As mentioned earlier, we first note that, under these distributions, a large fraction -- between $50\%$ and $80\%$ --- of the posterior distributions for the component black holes of GW190521 do not admit ancestors. In other words, under these spin constraints we obtain respective values of $p(\rm anc | d)$ of only $0.5$ and $0.2$}. This arises from the fact that such restricted ancestors can only yield final spins respectively limited to the ranges of \nncor{$a_{\rm child} \in [0.25,0.74]$} and $a_{\rm child} \in [0.38,0.69]$, while, under the priors imposed by our input analyses, the GW190521 spin magnitudes are fairly unconstrained.\textcolor{black}{ A similar situation occurs if we further restrict the ancestral mass-ratio to $q<2$. Doing so, we obtain respective values of $p(\rm anc | d) = 0.18$ and $0.1$.}\\

        With this, if GW190521 was quasi-circular, we obtain small values of \textcolor{black}{$p_{2g}$ in a Globular cluster, namely $p_{2g} =0.01$ for ancestral spin magnitudes below $0.2$ and $p_{2g} = 0.02$} for non-spinning ancestors. For a Nuclear Star Cluster both of these raise to \textcolor{black}{$p_{2g} = 0.3$}, which is nevertheless qualitatively consistent with the generic-spin case. \textcolor{black}{Further imposing ancestral mass-ratios $q<2$ leads to values of $p_{2g}=0.01$ under both spin restrictions for a Globular Cluster and $p_{2g}<0.2$} for a Nuclear Star Cluster. Second, if GW190521 was eccentric, we still obtain a practically null $p_{2g}$ for a Globular Cluster and $p_{2g} < 0.02$ for a Nuclear Star Cluster for both null and small spin magnitudes below 0.2, regardless the ancestral mass-ratio limits.\\

        As a summary, for the primary component of GW190521, \textcolor{black}{and under the analyses we use as input} \textcolor{black}{(and their corresponding priors)} we obtain four main conclusions:

        \begin{enumerate}
            \item If GW190521 was eccentric and we allow for generic ancestral spins, the chances that its primary BH is of second generation are highly suppressed. \textcolor{black}{We obtain $p_{2g} \leq 0.01$ for Globular Clusters and  $p_{2g} \simeq 0.1$ for the remaining environments, except for AGNs.} 
            
            \item If GW190521 happened in a Globular Cluster, its primary BH \textcolor{black}{only has a probability $p_{2g} > 0.1$ of being of second generation} if GW190521 was quasi-circular and its ancestors had aligned spins. \cor{The latter, however, is highly unlikely as aligned-spins are characteristic of isolated formation scenarios, with hierarchical formation channels within dense environments involving randomly oriented spins}. Still, in such a case, we obtain small values \textcolor{black}{$p_{2g} = 0.22 (0.14)$} if we restrict the ancestral mass ratio to $q\leq 2 (6)$. \textcolor{black}{For precessing ancestors, restricting the ancestral spins to magnitudes below $0.2$, does raise $p_{2g}$ to values of $\sim 10^{-2}$.}
            \item We only obtain $p_{2g} > 0.5$ if GW190521 was \textcolor{black}{quasi-circular} and hosted by a Nuclear Star Cluster, an Elliptical Galaxy or a Milky-Way-like Galaxy, provided that its ancestors had aligned spins. For precessing ancestors, we obtain $p_{2g}$ around 0.5 for Nuclear Star Clusters and Elliptic Galaxies. \textcolor{black}{For the latter two cases, we note that we obtain $p_{2g} \sim 0.7$ if we restrict to ancestral mass ratios  $q<2$ and generic spin magnitudes}.
            %\item \ncor{In addition, restricting the ancestral spin magnitudes to either null and small values, and considering a Globular Cluster host, we obtain a respective \nnncor{$p_{2g}$ values} of $0.1$ $(0.06)$, provided GW190521 was quasi-circular. For an Nuclear Star Cluster, we obtain $p_{2g} = 0.6$, similar to the generic spin case. }
            \item \cor{All in all, our results suggest that, \textcolor{black}{under the reference analyses of GW190521 we consider,} a hierarchical origin is best explained through a preceeding merger of ancestral BHs with randomly oriented spins in a very high $v_{esc}$ environment like an AGN or, more marginally, a Nuclear-Star Cluster.}
        \end{enumerate}

    \subsubsection{Secondary component of GW190521}

        For the secondary component of GW190521, we obtain results with a similar qualitative behaviour. The main quantitative difference resides in that $p_{2g}$ is much larger in the infinite escape velocity limit, due to the lower mass estimates for its ancestors. In particular, we always obtain $p_{2g} > 0.5$ for all hosts except for a Globular Cluster. For the latter, we obtain again $p_{2g} \simeq 0$  in all cases, except when we consider aligned-spin ancestors together with a quasi-circular scenario for GW190521, for which we obtain $p_{2g} \simeq 0.3$.

            \subsubsection{On the suitability of $p(\theta_{\rm anc})$ distributions as actual ancestral Bayesian priors given gravitational-wave data}\textcolor{black}{Importantly, we note that some of the restricted ancestral distributions discussed above lead to low values of $p_{2g}$ mostly due to the low values of $p(\rm anc | d)$ these yield given our input analyses. For instance, restricting the ancestors to $q<2$ and small spin magnitudes, and under the quasi-circular scenario, one obtains $p(\rm 2g | anc) = 0.95$ for a Nuclear Star Cluster, which is then reduced to $p_{2g}=0.17$ due to the fact that we obtain $p(\rm anc |d) = 0.18$. \footnote{We note, nevertheless, that for Globular Clusters we obtain $p(\rm 2g | anc) < 0.06$ in all cases where the ancestors are allowed to precess.} As stressed in our methods section, \textit{the fact that a given choice of $p(\theta_{anc})$ yields low values of $p(\rm anc | d)$ for our particular input analyses, does not necessarily mean that such hypotheses (null or small ancestral spins) provide a bad model for the gravitational-wave data}. Instead, it means that under the priors imposed on the ``child'' parameters by our input analyses, a large fraction of our input posterior distributions for such parameters do not admit ancestors within such ancestral population. Conversely, it follows that the imposition of such ancestral priors -- and those induced on the children parameters -- in a ``forward'' analysis like that in \cite{Mahapatra2024} would yield posterior spin distributions that would rail against the corresponding priors, ignoring a large amount of component spin parameters consistent with the data. Nevertheless, such an ancestral prior may actually pose a better model for the gravitational-wave data \footnote{In which one would trivially have $p(\rm anc | d)=1$.}, leading to a larger Bayesian evidence. This would be the case if the induced prior on the children parameters favours large-likelihood regions of the parameter space more than the rather agnostic uniform priors imposed by the input analyses we consider in this work.  As an example, we can roughly estimate the ratio of the Bayesian evidences between the non-spinning ancestral hypothesis for the primary BH ancestors and that imposed by the LVK  \footnote{We are ignoring the prior induced in the child masses, focusing only on that imposed on the spin magnitudes. Therefore the result of this calculation should be taken with caution.} through the prior re-weighting  technique as}

        \begin{equation}
            \frac{{\cal{Z}}^{a_i=0}_{\rm anc}}{{\cal{Z}}_{\rm LVK}} = \int d a_1 \frac{p^{a_i=0}_{\rm anc}(a_1)}{p_{\rm LVK}(a_1)} p(a_1 | d).
        \end{equation}

        \textcolor{black}{Above, the term  $p(a_1 | d)$ denotes the marginalized posterior distribution for the primary-spin magnitude of GW190521 under the LVK analysis while the terms $p^{a_i=0}_{\rm anc}(a_1)$ and $p_{\rm LVK}(a_1)$ respectively denote the prior on the spin magnitude of the primary black hole of GW190521 induced by the restriction of the ancestral black-holes being non-spinning and that imposed by the LVK, which is flat. This yields an evidence ratio of 1.2, indicating that both models are equally supported by the data. If instead, we consider the  eccentric posterior distributions of \cite{RomeroShaw2020_ecc_apjl}, we obtain an evidence ratio of 0.07, which disfavours  rather strongly the non-spinning ancestral hypothesis \textit{given the gravitational-wave data}}.\\

\section{Conclusions}
\label{SEC: Conclusion}

    We have described a Bayesian and self-consistent framework to obtain \cor{posterior distributions for the} masses and spins of the putative ancestors of the component merging black observed in gravitational-wave events, together with their corresponding ``birth kicks''. We do this by approximately solving the ``inverse final value'' problem: \cor{given posterior samples for the ``child'' black holes and given \textcolor{black}{proposal distributions $p(\theta_{\rm anc})$} for their ancestors, we obtain posterior distributions for the latter}. Such estimates are of particular relevance in the context of BBH observations involving component masses populating the PISN gap which. In particular, these allow us to assess \jc{the} viability of such BHs as products of previous mergers as a function of the properties of the host environment. We have applied this framework to \textcolor{black}{two existing analyses of the} \jc{mysterious} GW190521 event, whose primary black hole populates the PISN gap. First, to \textcolor{black}{the original study carried out by the LVK \cite{GW190521D}, which assumes GW190521 is a quasi-circular precessing merger. Second, to that by \cite{RomeroShaw2020_ecc_apjl}, which assumes GW190521 is a non-precessing eccentric merger. We have done this for three different distributions $p(\theta_{\rm anc})$ for the ancestors' parameters, also testing different restrictions in terms of the ancestral spins and mass ratios.} Next, we have estimated the \jc{probability} that the component black holes of GW190521 are the result of a merger of stellar-origin black holes, as a function of the host environment. \\ 
    
    Our estimates of the ancestor properties and the corresponding birth kicks can strongly depend on the combination of the scenario for GW190521 (eccentric or quasi-circular) and on the choice of $p(\theta_{\rm anc})$. Far from denoting lack of robustness, this underlines the potential of our our method to select between ancestral scenarios, specially if the host environment can be identified by complementary observations. Therefore, our method can drive highly relevant astrophysical conclusions, both about the properties of host environments and on the viability and existence of hierarchical BH formation channels. For instance, our results show that if GW190521 was eccentric, its primary component has less than $10\%$ chance of being born from the merger of stellar-born black-holes; with null chances if the host was a Globular Cluster. Moreover, \textcolor{black}{according to the two input analyses we consider \textcolor{black}{and their assumed priors}, a Globular Cluster is a viable host (with $p_{2g} > 0.1$) only} if GW190521 is a quasi-circular merger and the ancestors of its primary BH had aligned spins. \jc{Our results are broadly consistent with those by \cite{Kimball2021}, which concluded that the primary BH of GW190521 has a strong chance to be a second-generation BH if its host escape velocity exceeds $100$ km/s. In addition, \textit{our results support that GW190521 originated in an AGN \cite{Ford2022}, especially if it was eccentric}. On the one hand, the escape velocity of AGNs largely exceed the recoil velocities of BBH remnants. On the other hand, orbital eccentricity in GW190521, together with the large mass ratio and small spins needed to produce its components are also characteristic of BBHs in AGNs \cite{Ford2022}}. \cor{Finally, if GW190521 was quasi-circular, a Nuclear-Star origin is also possible, with $p_{2g}  \sim ( 0.4, 0.8) $.}\\

    We note that our framework, inspired by \cite{Barrera2022_masses,Barrera_spins}, is not the first one used to look into ancestral black holes. An alternative consists on simulating primary black-hole merger populations to then derive the properties of the second-generation (or further) remnant black-holes retained in selected environments \cite{Gerosa2019_escape_multiple_gen,Mould2022_DL_hierarchical_pop,Gerosa2021_heavyslow,Baibhav2021_LIGO_parents_21g_0412}. Finally, the properties of such remnants are contrasted with LIGO-Virgo observations to check for consistency. These techniques, however, do naturally fail to capture all the possible range of ancestral parameters due to, e.g., the restrictions imposed by the escape velocity of the host environment. \cor{In contrast, given a wide enough prior, our framework allows us to find posterior distributions covering all potential ancestral BBHs, which can then be flexibly re-weighted or restricted to alternative priors, as we do here with the characteristic escape velocity of certain environments.}

    Our work is limited by the following main aspects. First, we have obtained final masses, spins and kicks through the \texttt{NRSur7dq4Remnant} model. While this model provides accurate estimates for precessing BBHs, this also forces us to limit the mass ratio of the ancestor BBHs to $q\leq 6$, which in turn imposes constraints not only on the possible final spins, but also on the probability distribution of the kick for a given final spin. Second, while we have considered the case where GW190521 is an eccentric binary, we have worked under the assumption that the ancestor BHs described quasi-circular ones. Relaxing this assumption shall lead to significantly different probability distributions for the final mass and spins of the corresponding remnants and, possibly, to different astrophysical conclusions. We note that while remnant models for eccentric BBHs are being developed \cite{EccSur_Boschini2023}, these still lack the kick information, which is key for our work. \jc{Third, we have not considered a scenario for GW190521 combining both precession and eccentricity, as such study would require of fast-to-evaluate waveform models including both effects, which do not yet exist. We note that \cite{Gayathri2022_ecc_natastro} performed a study combining both effects, finding a good fit to GW190521, using numerical relativity waveforms. However, this study only reports a single point-like parameter estimate for a best-fit numerical simulation with fixed mass ratio, spins and eccentricity, making it unrealistic to estimate how our results would look under such scenario} \footnote{We note also that while \cite{Toni_ecc} analysed GW190521 under an eccentric non-precessing scenario, they do not report the spin magnitude estimates, which are key in our study.}. Fourth, it has been shown that environmental effects can induce changes in the GW emission around the merger and ringdown stages of black-hole mergers \cite{Leong2023}, which can in turn modify the kick with respect to the vacuum case \cite{Zhang2023}.\footnote{We note, however, that the mentioned studies consider somewhat mock environments consisting \jc{of} scalar-field bubbles. Therefore, while the results reported should qualitatively hold for more realistic environments, quantitative ones such as the actual impact on the kick will most likely not apply.} Finally, while GW190521 is the most confident BBH detection to date with component masses within the PISN gap, less confident events like GW200220 \cite{abbott2021gwtc3} or GW190426 \cite{GWTC2.1} pose also suitable candidates for displaying second-generation BHs. In forthcoming studies, we aim to expand our study to further final-parameter models, suitable ancestral BH priors and a larger suite of BBH detections, as these are provided by current and future observing runs.\\
    
    Our work sets the ground for full parameter inference of ancestral BBHs, highlighting the crucial role of birth kick estimates and how these shall allow to select between combinations of potential ancestor populations and host environments. This, in turn, can help to extract crucial astrophysical information such as the properties of host environments of BBH mergers and the viability of hierarchical formation channels as those giving origin to the high-mass BHs observed by GW detectors. The rather poor accuracy of current spin-magnitude estimates in BBH events significantly limits that of the putative birth kicks, preventing stronger astrophysical conclusions. This situation shall be much improved as the GW detector network progressively improves its sensitivity, particularly with the arrival of third-generation detectors like Einstein Telescope \cite{ET1,ET2} or Cosmic Explorer \cite{CE,CE2}.\\

    Finally, we note that while finalizing this paper, we learned that \cite{Mahapatra2024} have also looked into the problem of identifying the parents of GW190521 assuming hierarchical merger scenario, using a Bayesian framework. As discussed throughout the text, the main difference between our ``backwards'' formalism and the ``forward'' one in \cite{Mahapatra2024} resides in that we condition our results on the posterior distribution for the component BHs obtained by existing analyses of the GW signal. \textcolor{black}{Therefore we naturally inherit and keep intact both the priors imposed by such analyses on the children black-holes and the corresponding posteriors, to then posterior parameter estimates for the ancestors. Instead, in their ``forward'' method, \cite{Mahapatra2024} \textit{impose priors} on the ancestral parameters to then simultaneously obtain \textit{their own posterior distributions on both the components of GW190521 and the corresponding ancestors}, i.e., they re-analyze the data under different priors. In simpler words, we aim to answer the question ``\textit{Given the parameters of the GW190521 components inferred by an existing analysis, can we infer the parameters of their putative ancestors?}''. In contrast, \cite{Mahapatra2024} answer the question ``\textit{Given gravitational-wave data, and assuming it was produced by the merger of second-generation black holes, what are the parameters of both the merger components and their ancestors?}''.} We note that both works, nevertheless, obtain consistent results when assuming that GW190521 was quasi-circular, which \cite{Mahapatra2024} restricts to. 

\section{acknowledgments}

    We thank Isobel Romero-Shaw for kindly sharing the posterior distributions for GW190521 under the eccentric scenario and for many useful comments on the manuscript. We thank Thomas Dent, Michella Mapelli, Parthapratim Mahapatra, Christopher Berry, Imre Bartos, Ester Ruiz-Morales and Davide Gerosa for useful comments. We deeply thank Matthew Mould for detecting an inconsistency in the derivation of our formalism presented in our first submission and for enlightening discussions. We also thank Kiki Wing Lo and Isabel Vazquez-Martinez for useful discussions during the early stages of this study. JCB is funded by a fellowship from ``la Caixa'' Foundation (ID100010474) and from the European Union's Horizon2020 research and innovation programme under the Marie Skodowska-Curie grant agreement No 847648. The fellowship code is LCF/BQ/PI20/11760016. JCB is also supported by the research grant PID2020-118635GB-I00 from the Spain-Ministerio de Ciencia e Innovaci\'{o}n and by the European Horizon Europe staff exchange (SE) programme HORIZON-MSCA2021-SE-01 Grant No. NewFunFiCO-101086251. We acknowledge the use of IUCAA LDG cluster Sarathi for the computational/numerical work. The authors acknowledge computational resources provided by the CIT cluster of the LIGO Laboratory and supported by National Science Foundation Grants PHY-0757058 and PHY0823459; and the support of the NSF CIT cluster for the provision of computational resources for our parameter inference runs. This material is based upon work supported by NSF's LIGO Laboratory which is a major facility fully funded by the National Science Foundation. The analysed LIGO-Virgo data and the corresponding power spectral densities, in their strain versions, are publicly available at the online Gravitational-Wave Open Science Center~\cite{SoftwareX,OpenDataArxiv}. This research has made use of data or software obtained from the Gravitational Wave Open Science Center (gwosc.org), a service of LIGO Laboratory, the LIGO Scientific Collaboration, the Virgo Collaboration, and KAGRA. LIGO Laboratory and Advanced LIGO are funded by the United States National Science Foundation (NSF) as well as the Science and Technology Facilities Council (STFC) of the United Kingdom, the Max-Planck-Society (MPS), and the State of Niedersachsen/Germany for support of the construction of Advanced LIGO and construction and operation of the GEO600 detector. Additional support for Advanced LIGO was provided by the Australian Research Council. Virgo is funded, through the European Gravitational Observatory (EGO), by the French Centre National de Recherche Scientifique (CNRS), the Italian Istituto Nazionale di Fisica Nucleare (INFN) and the Dutch Nikhef, with contributions by institutions from Belgium, Germany, Greece, Hungary, Ireland, Japan, Monaco, Poland, Portugal, Spain. KAGRA is supported by Ministry of Education, Culture, Sports, Science and Technology (MEXT), Japan Society for the Promotion of Science (JSPS) in Japan; National Research Foundation (NRF) and Ministry of Science and ICT (MSIT) in Korea; Academia Sinica (AS) and National Science and Technology Council (NSTC) in Taiwan. This manuscript has LIGO DCC number P2400073.

\bibliography{IMBBH.bib,psi4_observation}{}

\begin{thebibliography}{}
\expandafter\ifx\csname natexlab\endcsname\relax\def\natexlab#1{#1}\fi
\providecommand{\url}[1]{\href{#1}{#1}}
\providecommand{\dodoi}[1]{doi:~\href{http://doi.org/#1}{\nolinkurl{#1}}}
\providecommand{\doeprint}[1]{\href{http://ascl.net/#1}{\nolinkurl{http://ascl.net/#1}}}
\providecommand{\doarXiv}[1]{\href{https://arxiv.org/abs/#1}{\nolinkurl{https://arxiv.org/abs/#1}}}

\bibitem[{Aasi {et~al.}(2015)}]{AdvancedLIGOREF}
Aasi, J., {et~al.} 2015, Classical and Quantum Gravity, 32, 074001,
  \dodoi{10.1088/0264-9381/32/7/074001}

\bibitem[{Abbott {et~al.}(2020{\natexlab{a}})}]{GW190521D}
Abbott, {et~al.} 2020{\natexlab{a}}, Physical Review Letters, 125,
  \dodoi{10.1103/PhysRevLett.125.101102}

\bibitem[{Abbott {et~al.}(2020{\natexlab{b}})}]{GW190521I}
Abbott, B., {et~al.} 2020{\natexlab{b}}, Astrophys. J. Lett., 900, L13,
  \dodoi{doi.org/10.3847/2041-8213/aba493}

\bibitem[{Abbott {et~al.}(2017)}]{CE2}
Abbott, B.~P., {et~al.} 2017, Classical and Quantum Gravity, 34, 044001,
  \dodoi{10.1088/1361-6382/aa51f4}

\bibitem[{Abbott {et~al.}(2021{\natexlab{a}})Abbott, Abbott, Acernese, Ackley,
  Adams, Adhikari, Adhikari, Adya, Affeldt, Agarwal,
  {et~al.}}]{abbott2021gwtc3}
Abbott, R., Abbott, T., Acernese, F., {et~al.} 2021{\natexlab{a}}, arXiv
  preprint arXiv:2111.03606

\bibitem[{Abbott {et~al.}(2021{\natexlab{b}})}]{GWTC3-pop}
Abbott, R., {et~al.} 2021{\natexlab{b}}, The population of merging compact
  binaries inferred using gravitational waves through GWTC-3,  arXiv,
  \dodoi{10.48550/ARXIV.2111.03634}

\bibitem[{Abbott {et~al.}(2021{\natexlab{c}})}]{SoftwareX}
---. 2021{\natexlab{c}}, {SoftwareX}, 13, 100658,
  \dodoi{10.1016/j.softx.2021.100658}

\bibitem[{Acernese {et~al.}(2015)}]{TheVirgo:2014hva}
Acernese, F., {et~al.} 2015, Class. Quant. Grav., 32, 024001,
  \dodoi{10.1088/0264-9381/32/2/024001}

\bibitem[{Akutsu {et~al.}(2020)}]{akutsu2020overview}
Akutsu, T., {et~al.} 2020, Overview of {KAGRA}: Detector design and
  construction history.
\newblock \doarXiv{2005.05574}

\bibitem[{Anagnostou {et~al.}(2022)Anagnostou, Trenti, \&
  Melatos}]{Anagnostou2022}
Anagnostou, O., Trenti, M., \& Melatos, A. 2022, The Astrophysical Journal,
  941, 4, \dodoi{10.3847/1538-4357/ac9d95}

\bibitem[{Baibhav {et~al.}(2021)Baibhav, Berti, Gerosa, Mould, \&
  Wong}]{Baibhav2021_LIGO_parents_21g_0412}
Baibhav, V., Berti, E., Gerosa, D., Mould, M., \& Wong, K.~W. 2021, Physical
  Review D, 104, \dodoi{10.1103/physrevd.104.084002}

\bibitem[{Barrera \& Bartos(2022)}]{Barrera2022_masses}
Barrera, O., \& Bartos, I. 2022, The Astrophysical Journal Letters, 929, L1,
  \dodoi{10.3847/2041-8213/ac5f47}

\bibitem[{Barrera \& Bartos(2023)}]{Barrera_spins}
---. 2023, Ancestral Spin Information in Gravitational Waves from Black Hole
  Mergers

\bibitem[{Baumgardt \& Hilker(2018)}]{Baumgardt2018}
Baumgardt, H., \& Hilker, M. 2018, Monthly Notices of the Royal Astronomical
  Society, 478, 1520, \dodoi{10.1093/mnras/sty1057}

\bibitem[{Belczynski {et~al.}(2016)Belczynski, Holz, Bulik, \&
  O’Shaughnessy}]{Belczynski2016}
Belczynski, K., Holz, D.~E., Bulik, T., \& O’Shaughnessy, R. 2016, Nature,
  534, 512–515, \dodoi{10.1038/nature18322}

\bibitem[{Boschini {et~al.}(2023)Boschini, Gerosa, Varma, Armaza, Boyle,
  Bonilla, Ceja, Chen, Deppe, Giesler, Kidder, Kumar, Lara, Long, Ma, Mitman,
  Nee, Pfeiffer, Ramos-Buades, Scheel, Vu, \& Yoo}]{EccSur_Boschini2023}
Boschini, M., Gerosa, D., Varma, V., {et~al.} 2023, Physical Review D, 108,
  \dodoi{10.1103/physrevd.108.084015}

\bibitem[{Br\"{u}gmann {et~al.}(2008)Br\"{u}gmann, González, Hannam, Husa, \&
  Sperhake}]{Brgmann2008}
Br\"{u}gmann, B., González, J.~A., Hannam, M., Husa, S., \& Sperhake, U. 2008,
  Physical Review D, 77, \dodoi{10.1103/physrevd.77.124047}

\bibitem[{Bustillo {et~al.}(2022)Bustillo, Leong, \& Chandra}]{GW190412_recoil}
Bustillo, J.~C., Leong, S. H.~W., \& Chandra, K. 2022, GW190412: measuring a
  black-hole recoil direction through higher-order gravitational-wave modes

\bibitem[{Bustillo {et~al.}(2021{\natexlab{a}})Bustillo, Sanchis-Gual,
  Torres-Forné, \& Font}]{Bustillo2021_Headon}
Bustillo, J.~C., Sanchis-Gual, N., Torres-Forné, A., \& Font, J.~A.
  2021{\natexlab{a}}, Physical Review Letters, 126,
  \dodoi{10.1103/physrevlett.126.201101}

\bibitem[{Bustillo {et~al.}(2021{\natexlab{b}})Bustillo, Sanchis-Gual,
  Torres-Forn{\'{e}}, Font, Vajpeyi, Smith, Herdeiro, Radu, \& Leong}]{Proca}
Bustillo, J.~C., Sanchis-Gual, N., Torres-Forn{\'{e}}, A., {et~al.}
  2021{\natexlab{b}}, Physical Review Letters, 126,
  \dodoi{10.1103/physrevlett.126.081101}

\bibitem[{Calder{\'o}n~Bustillo {et~al.}(2018)Calder{\'o}n~Bustillo, Clark,
  Laguna, \& Shoemaker}]{CalderonBustillo:2018zuq}
Calder{\'o}n~Bustillo, J., Clark, J.~A., Laguna, P., \& Shoemaker, D. 2018,
  Phys. Rev. Lett., 121, 191102, \dodoi{10.1103/PhysRevLett.121.191102}

\bibitem[{Calderón~Bustillo {et~al.}(2023)Calderón~Bustillo, Sanchis-Gual,
  Leong, Chandra, Torres-Forné, Font, Herdeiro, Radu, Wong, \&
  Li}]{Observations_proca}
Calderón~Bustillo, J., Sanchis-Gual, N., Leong, S.~H., {et~al.} 2023, Physical
  Review D, 108, \dodoi{10.1103/physrevd.108.123020}

\bibitem[{Cao \& Han(2017)}]{Cao2017}
Cao, Z., \& Han, W.-B. 2017, Physical Review D, 96,
  \dodoi{10.1103/physrevd.96.044028}

\bibitem[{Chandra {et~al.}(2023)Chandra, Pai, Leong, \&
  Bustillo}]{koustav_2309.01683}
Chandra, K., Pai, A., Leong, S. H.~W., \& Bustillo, J.~C. 2023, Impact of
  Bayesian Priors on the Inferred Masses of Quasi-Circular Intermediate-Mass
  Black Hole Binaries

\bibitem[{Collaboration \& the Virgo~Collaboration(2021)}]{GWTC2.1}
Collaboration, T. L.~S., \& the Virgo~Collaboration. 2021, GWTC-2.1: Deep
  Extended Catalog of Compact Binary Coalescences Observed by LIGO and Virgo
  During the First Half of the Third Observing Run

\bibitem[{Collaboration {et~al.}(2023)Collaboration, the Virgo~Collaboration,
  \& the KAGRA~Collaboration}]{OpenDataArxiv}
Collaboration, T. L.~S., the Virgo~Collaboration, \& the KAGRA~Collaboration.
  2023, The Astrophysical Journal Supplement Series, 2,
  \dodoi{10.1103/PhysRevD.100.064064}

\bibitem[{Costa {et~al.}(2022)Costa, Ballone, Mapelli, \& Bressan}]{Costa2022}
Costa, G., Ballone, A., Mapelli, M., \& Bressan, A. 2022, Monthly Notices of
  the Royal Astronomical Society, 516, 1072–1080,
  \dodoi{10.1093/mnras/stac2222}

\bibitem[{Estell{\'{e}}s {et~al.}(2022)Estell{\'{e}}s, Husa, Colleoni,
  Mateu-Lucena, de~Lluc~Planas, Garc{\'{\i}}a-Quir{\'{o}}s, Keitel,
  Ramos-Buades, Mehta, Buonanno, \& Ossokine}]{Estells2022_GW190521}
Estell{\'{e}}s, H., Husa, S., Colleoni, M., {et~al.} 2022, The Astrophysical
  Journal, 924, 79, \dodoi{10.3847/1538-4357/ac33a0}

\bibitem[{Farmer {et~al.}(2019)Farmer, Renzo, de~Mink, Marchant, \&
  Justham}]{Farmer:2019jed}
Farmer, R., Renzo, M., de~Mink, S.~E., Marchant, P., \& Justham, S. 2019,
  \dodoi{10.3847/1538-4357/ab518b}

\bibitem[{Farr {et~al.}(2017)Farr, Stevenson, Coleman~Miller, Mandel, Farr, \&
  Vecchio}]{Farr:2017uvj}
Farr, W.~M., Stevenson, S., Coleman~Miller, M., {et~al.} 2017, Nature, 548,
  426, \dodoi{10.1038/nature23453}

\bibitem[{Finn(1992)}]{Finn1992}
Finn, L.~S. 1992, Physical Review D, 46, 5236, \dodoi{10.1103/physrevd.46.5236}

\bibitem[{Fishbach \& Holz(2020)}]{Fishbach_2020}
Fishbach, M., \& Holz, D.~E. 2020, The Astrophysical Journal Letters, 904, L26,
  \dodoi{10.3847/2041-8213/abc827}

\bibitem[{Ford \& McKernan(2022)}]{Ford2022}
Ford, K. E.~S., \& McKernan, B. 2022, Monthly Notices of the Royal Astronomical
  Society, 517, 5827–5834, \dodoi{10.1093/mnras/stac2861}

\bibitem[{Fuller \& Ma(2019)}]{Fuller2019}
Fuller, J., \& Ma, L. 2019, The Astrophysical Journal Letters, 881, L1,
  \dodoi{10.3847/2041-8213/ab339b}

\bibitem[{Gamba {et~al.}(2022)Gamba, Breschi, Carullo, Albanesi, Rettegno,
  Bernuzzi, \& Nagar}]{Gamba2022_ecc_natastro}
Gamba, R., Breschi, M., Carullo, G., {et~al.} 2022, Nature Astronomy, 7, 11,
  \dodoi{10.1038/s41550-022-01813-w}

\bibitem[{Gayathri {et~al.}(2022)Gayathri, Healy, Lange, O'Brien,
  Szczepa{\'{n}}czyk, Bartos, Campanelli, Klimenko, Lousto, \&
  O'Shaughnessy}]{Gayathri2022_ecc_natastro}
Gayathri, V., Healy, J., Lange, J., {et~al.} 2022, Nature Astronomy, 6, 344,
  \dodoi{10.1038/s41550-021-01568-w}

\bibitem[{Gerosa \& Berti(2019)}]{Gerosa2019_escape_multiple_gen}
Gerosa, D., \& Berti, E. 2019, Physical Review D, 100,
  \dodoi{10.1103/physrevd.100.041301}

\bibitem[{Gerosa \& Fishbach(2021)}]{Gerosa2021_Fischbach_review}
Gerosa, D., \& Fishbach, M. 2021, Nature Astronomy, 5, 749–760,
  \dodoi{10.1038/s41550-021-01398-w}

\bibitem[{Gerosa {et~al.}(2021)Gerosa, Giacobbo, \&
  Vecchio}]{Gerosa2021_heavyslow}
Gerosa, D., Giacobbo, N., \& Vecchio, A. 2021, The Astrophysical Journal, 915,
  56, \dodoi{10.3847/1538-4357/ac00bb}

\bibitem[{Giacobbo \& Mapelli(2018)}]{Giacobbo_2018}
Giacobbo, N., \& Mapelli, M. 2018, Monthly Notices of the Royal Astronomical
  Society, 480, 2011–2030, \dodoi{10.1093/mnras/sty1999}

\bibitem[{Gonzalez {et~al.}(2007)Gonzalez, Sperhake, Bruegmann, Hannam, \&
  Husa}]{Gonzalez:2006md}
Gonzalez, J.~A., Sperhake, U., Bruegmann, B., Hannam, M., \& Husa, S. 2007,
  Phys. Rev. Lett., 98, 091101, \dodoi{10.1103/PhysRevLett.98.091101}

\bibitem[{Healy {et~al.}(2014)Healy, Lousto, \&
  Zlochower}]{Healy2014_alignedspinkick}
Healy, J., Lousto, C.~O., \& Zlochower, Y. 2014, Physical Review D, 90,
  \dodoi{10.1103/physrevd.90.104004}

\bibitem[{Heger {et~al.}(2003)Heger, Fryer, Woosley, Langer, \&
  Hartmann}]{Heger:2002by}
Heger, A., Fryer, C.~L., Woosley, S.~E., Langer, N., \& Hartmann, D.~H. 2003,
  Astrophys. J., 591, 288, \dodoi{10.1086/375341}

\bibitem[{Hild {et~al.}(2010)}]{ET2}
Hild, S., {et~al.} 2010, \dodoi{10.1088/0264-9381/28/9/094013}

\bibitem[{Kimball {et~al.}(2021{\natexlab{a}})Kimball, Talbot, Berry, Zevin,
  Thrane, Kalogera, Buscicchio, Carney, Dent, Middleton, Payne, Veitch, \&
  Williams}]{Kimball_2021}
Kimball, C., Talbot, C., Berry, C. P.~L., {et~al.} 2021{\natexlab{a}}, The
  Astrophysical Journal Letters, 915, L35, \dodoi{10.3847/2041-8213/ac0aef}

\bibitem[{Kimball {et~al.}(2021{\natexlab{b}})Kimball, Talbot, Berry, Zevin,
  Thrane, Kalogera, Buscicchio, Carney, Dent, Middleton, Payne, Veitch, \&
  Williams}]{Kimball2021}
---. 2021{\natexlab{b}}, The Astrophysical Journal Letters, 915, L35,
  \dodoi{10.3847/2041-8213/ac0aef}

\bibitem[{Leong {et~al.}(2023)Leong, Calderón~Bustillo, Gracia-Linares, \&
  Laguna}]{Leong2023}
Leong, S.~H., Calderón~Bustillo, J., Gracia-Linares, M., \& Laguna, P. 2023,
  Physical Review D, 108, \dodoi{10.1103/physrevd.108.124079}

\bibitem[{Liu {et~al.}(2019)Liu, Cao, \& Shao}]{Liu_SEOBNRE}
Liu, X., Cao, Z., \& Shao, L. 2019, \dodoi{10.1103/PhysRevD.101.044049}

\bibitem[{Mahapatra {et~al.}(2024)Mahapatra, Chattopadhyay, Gupta, Antonini,
  Favata, Sathyaprakash, \& Arun}]{Mahapatra2024}
Mahapatra, P., Chattopadhyay, D., Gupta, A., {et~al.} 2024, Reconstructing the
  genealogy of LIGO-Virgo black holes

\bibitem[{Mahapatra {et~al.}(2021)Mahapatra, Gupta, Favata, Arun, \&
  Sathyaprakash}]{Measured_kick_magnitude_GW190814}
Mahapatra, P., Gupta, A., Favata, M., Arun, K.~G., \& Sathyaprakash, B.~S.
  2021, The Astrophysical Journal Letters, 918, L31,
  \dodoi{10.3847/2041-8213/ac20db}

\bibitem[{Mahapatra {et~al.}(2022)Mahapatra, Gupta, Favata, Arun, \&
  Sathyaprakash}]{2209.05766_mahapatra}
---. 2022, Black hole hierarchical growth efficiency and mass spectrum
  predictions

\bibitem[{Mandel \& de Mink(2016)}]{Mandel_2016}
Mandel, I., \& de Mink, S.~E. 2016, Monthly Notices of the Royal Astronomical
  Society, 458, 2634–2647, \dodoi{10.1093/mnras/stw379}

\bibitem[{Mapelli {et~al.}(2020)Mapelli, Spera, Montanari, Limongi, Chieffi,
  Giacobbo, Bressan, \& Bouffanais}]{Mapelli2020}
Mapelli, M., Spera, M., Montanari, E., {et~al.} 2020, The Astrophysical
  Journal, 888, 76, \dodoi{10.3847/1538-4357/ab584d}

\bibitem[{Mould {et~al.}(2022)Mould, Gerosa, \&
  Taylor}]{Mould2022_DL_hierarchical_pop}
Mould, M., Gerosa, D., \& Taylor, S.~R. 2022, Physical Review D, 106,
  \dodoi{10.1103/physrevd.106.103013}

\bibitem[{Natarajan \& Pringle(1998)}]{Natarajan1998}
Natarajan, P., \& Pringle, J.~E. 1998, The Astrophysical Journal, 506,
  L97–L100, \dodoi{10.1086/311658}

\bibitem[{Nitz \& Capano(2021)}]{GW190521_Nitz}
Nitz, A.~H., \& Capano, C.~D. 2021, The Astrophysical Journal, 907, L9,
  \dodoi{10.3847/2041-8213/abccc5}

\bibitem[{Palmese \& Conselice(2021)}]{Palmese2021}
Palmese, A., \& Conselice, C.~J. 2021, Physical Review Letters, 126,
  \dodoi{10.1103/physrevlett.126.181103}

\bibitem[{Payne {et~al.}(2019)Payne, Talbot, \& Thrane}]{Payne2019_rew}
Payne, E., Talbot, C., \& Thrane, E. 2019, Physical Review D, 100,
  \dodoi{10.1103/physrevd.100.123017}

\bibitem[{Portegies~Zwart \& McMillan(2000)}]{PortegiesZwart2000}
Portegies~Zwart, S.~F., \& McMillan, S. L.~W. 2000, The Astrophysical Journal,
  528, L17–L20, \dodoi{10.1086/312422}

\bibitem[{Punturo {et~al.}(2010)}]{ET1}
Punturo, M., {et~al.} 2010, Classical and Quantum Gravity, 27, 084007,
  \dodoi{10.1088/0264-9381/27/8/084007}

\bibitem[{Ramos-Buades {et~al.}(2023)Ramos-Buades, Buonanno, \&
  Gair}]{Toni_ecc}
Ramos-Buades, A., Buonanno, A., \& Gair, J. 2023, Bayesian inference of binary
  black holes with inspiral-merger-ringdown waveforms using two eccentric
  parameters

\bibitem[{Reitze {et~al.}(2019)Reitze, Adhikari, Ballmer, Barish, Barsotti,
  Billingsley, Brown, Chen, Coyne, Eisenstein, Evans, Fritschel, Hall,
  Lazzarini, Lovelace, Read, Sathyaprakash, Shoemaker, Smith, Torrie, Vitale,
  Weiss, Wipf, \& Zucker}]{CE}
Reitze, D., Adhikari, R.~X., Ballmer, S., {et~al.} 2019

\bibitem[{Rodriguez {et~al.}(2016)Rodriguez, Zevin, Pankow, Kalogera, \&
  Rasio}]{Rodriguez2016}
Rodriguez, C.~L., Zevin, M., Pankow, C., Kalogera, V., \& Rasio, F.~A. 2016,
  The Astrophysical Journal Letters, 832, L2,
  \dodoi{10.3847/2041-8205/832/1/l2}

\bibitem[{Romano \& Cornish(2017)}]{Romano2017}
Romano, J.~D., \& Cornish, N.~J. 2017, Living Reviews in Relativity, 20,
  \dodoi{10.1007/s41114-017-0004-1}

\bibitem[{Romero-Shaw {et~al.}(2020{\natexlab{a}})Romero-Shaw, Lasky, Thrane,
  \& Bustillo}]{RomeroShaw2020_ecc_apjl}
Romero-Shaw, I., Lasky, P.~D., Thrane, E., \& Bustillo, J.~C.
  2020{\natexlab{a}}, The Astrophysical Journal, 903, L5,
  \dodoi{10.3847/2041-8213/abbe26}

\bibitem[{Romero-Shaw {et~al.}(2020{\natexlab{b}})Romero-Shaw, Lasky, Thrane,
  \& Calderon~Bustillo}]{Isobel_ecc}
Romero-Shaw, I.~M., Lasky, P.~D., Thrane, E., \& Calderon~Bustillo, J.
  2020{\natexlab{b}}, {GW190521}: orbital eccentricity and signatures of
  dynamical formation in a binary black hole merger signal

\bibitem[{Santamaria {et~al.}(2010)}]{Santamaria:2010yb}
Santamaria, L., {et~al.} 2010, Phys. Rev., D82, 064016,
  \dodoi{10.1103/PhysRevD.82.064016}

\bibitem[{Schmidt {et~al.}(2012)Schmidt, Hannam, \& Husa}]{Schmidt:2012rh}
Schmidt, P., Hannam, M., \& Husa, S. 2012, Phys. Rev., D86, 104063,
  \dodoi{10.1103/PhysRevD.86.104063}

\bibitem[{Schmidt {et~al.}(2015)Schmidt, Ohme, \& Hannam}]{Schmidt:2014iyl}
Schmidt, P., Ohme, F., \& Hannam, M. 2015, Phys. Rev., D91, 024043,
  \dodoi{10.1103/PhysRevD.91.024043}

\bibitem[{Sigurdsson \& Hernquist(1993)}]{Sigurdsson1993}
Sigurdsson, S., \& Hernquist, L. 1993, Nature, 364, 423–425,
  \dodoi{10.1038/364423a0}

\bibitem[{Sperhake {et~al.}(2011)Sperhake, Berti, Cardoso, Pretorius, \&
  Yunes}]{Sperhake2011}
Sperhake, U., Berti, E., Cardoso, V., Pretorius, F., \& Yunes, N. 2011,
  Physical Review D, 83, \dodoi{10.1103/physrevd.83.024037}

\bibitem[{Talbot \& Thrane(2017)}]{Talbot2017}
Talbot, C., \& Thrane, E. 2017, Physical Review D, 96,
  \dodoi{10.1103/physrevd.96.023012}

\bibitem[{Tanikawa {et~al.}(2021)Tanikawa, Kinugawa, Yoshida, Hijikawa, \&
  Umeda}]{Tanikawa2021}
Tanikawa, A., Kinugawa, T., Yoshida, T., Hijikawa, K., \& Umeda, H. 2021,
  Monthly Notices of the Royal Astronomical Society, 505, 2170–2176,
  \dodoi{10.1093/mnras/stab1421}

\bibitem[{Thorne(1980)}]{Thorne:1980ru}
Thorne, K.~S. 1980, Rev. Mod. Phys., 52, 299, \dodoi{10.1103/RevModPhys.52.299}

\bibitem[{Tutukov \& YungelSon(1993)}]{Tutukov1993}
Tutukov, A.~V., \& YungelSon, L.~R. 1993, Monthly Notices of the Royal
  Astronomical Society, 260, 675–678, \dodoi{10.1093/mnras/260.3.675}

\bibitem[{Varma {et~al.}(2019{\natexlab{a}})Varma, Field, Scheel, Blackman,
  Gerosa, Stein, Kidder, \& Pfeiffer}]{NRSur7dq4}
Varma, V., Field, S.~E., Scheel, M.~A., {et~al.} 2019{\natexlab{a}}, Physical
  Review Research, 1, \dodoi{10.1103/physrevresearch.1.033015}

\bibitem[{Varma {et~al.}(2019{\natexlab{b}})Varma, Gerosa, Stein, Hébert, \&
  Zhang}]{Varma2019_surrogate_remnant}
Varma, V., Gerosa, D., Stein, L.~C., Hébert, F., \& Zhang, H.
  2019{\natexlab{b}}, Physical Review Letters, 122,
  \dodoi{10.1103/physrevlett.122.011101}

\bibitem[{Varma {et~al.}(2020)Varma, Isi, \& Biscoveanu}]{Varma2020_kickstudy}
Varma, V., Isi, M., \& Biscoveanu, S. 2020, Physical Review Letters, 124,
  \dodoi{10.1103/physrevlett.124.101104}

\bibitem[{Varma {et~al.}(2022)Varma, Biscoveanu, Islam, Shaik, Haster, Isi,
  Farr, Field, \& Vitale}]{Vijay_GWKick}
Varma, V., Biscoveanu, S., Islam, T., {et~al.} 2022, Evidence of large recoil
  velocity from a black hole merger signal

\bibitem[{Winch {et~al.}(2024)Winch, Vink, Higgins, \&
  Sabhahit}]{2401.17327_Winch}
Winch, E. R.~J., Vink, J.~S., Higgins, E.~R., \& Sabhahit, G.~N. 2024,
  Predicting the Heaviest Black Holes below the Pair Instability Gap

\bibitem[{Woosley \& Heger(2021)}]{Woosley2021}
Woosley, S.~E., \& Heger, A. 2021, The Astrophysical Journal Letters, 912, L31,
  \dodoi{10.3847/2041-8213/abf2c4}

\bibitem[{Yang {et~al.}(2019)Yang, Bartos, Gayathri, Ford, Haiman, Klimenko,
  Kocsis, Márka, Márka, McKernan, \& O’Shaughnessy}]{Yang2019}
Yang, Y., Bartos, I., Gayathri, V., {et~al.} 2019, Physical Review Letters,
  123, \dodoi{10.1103/physrevlett.123.181101}

\bibitem[{Zhang {et~al.}(2023)Zhang, Gracia-Linares, Laguna, Shoemaker, \&
  Liu}]{Zhang2023}
Zhang, Y.-P., Gracia-Linares, M., Laguna, P., Shoemaker, D., \& Liu, Y.-X.
  2023, Physical Review D, 107, \dodoi{10.1103/physrevd.107.044039}

\end{thebibliography}
\bibliographystyle{aasjournal}

\end{document}